\documentclass[12pt]{JHEP3}

\newcommand{\be}{\begin{equation}}
\newcommand{\ee}{\end{equation}}
\newcommand{\bea}{\begin{eqnarray}}
\newcommand{\eea}{\end{eqnarray}}
\newcommand{\barr}{\begin{array}}
\newcommand{\earr}{\end{array}}
\usepackage{amsmath}
\title{Second order transport from anomalies}
\author{  Sayantani Bhattacharyya, ${}^a$ Justin R. David {${}^b$} and  Somyadip
Thakur {${}^b$}.  \\
${}^a$ Physics Department, \\
Ramkrishna Mission Vivekananda University,\\
Belur Math, Howrah 711202, India. \\
${}^b$ Centre for High Energy Physics,
Indian Institute of Science,\\ C.V. Raman Avenue, Bangalore 560012, India. \\
\email{sayanta@rkmvu.ac.in, justin,somyadip@cts.iisc.ernet.in}\\
\\ 
}

\abstract{We study parity odd transport  at  second order in
derivative expansion for a non-conformal charged  fluid. 
We see that there are 27 parity odd transport coefficients, of which 
12 are non-vanishing in equilibrium. 
We use the equilibrium partition function method to 
express $7$ of these in terms of the anomaly, shear viscosity, charge diffusivity
 and thermodynamic functions.   
The remaining $5$ are constrained by $3$ relations which also involve the anomaly. 
We  derive Kubo formulae for $2$  of the transport coefficients and show
these agree with that derived from the equilibrium partition function. 
}
\preprint{}

\begin{document}

\section{Introduction}

Fluid dynamics is an effective description of near-equilibrium systems. Properties of a fluid system are always slowly varying compared to some intrinsic length scale,  for example a mean free path. 
This length scale is determined by the details of the underlying microscopic theory.  
The fundamental variables  of  fluid dynamics are local velocities $u^\mu(x)$, temperature $T(x)$, 
 and all the other conserved charges or their chemical potentials $\mu^a(x)$. The 
conservation equations for the stress tensor $T^{\mu\nu}$ and the other conserved currents 
$J^{a\mu}$ govern the time evolution of fluid dynamics. 

  The stress tensor  and the  conserved currents are related to the fluid variables 
  $\{ u^\mu, T, \mu^a \}$  via constitutive relations.  Since fluid systems are always slowly varying 
it is appropriate  to organize  the constitutive relations in terms of  a derivative expansion  of the 
fluid variables.   At every order in the  derivative expansion,  the independent terms  of the constitutive 
relation are constructed out of the independent derivatives of the fluid variables. 
 The  independent terms  in the 
 constitutive relation are multiplied by coefficients which are 
functions of temperature and chemical potentials. These coefficients are called transport coefficients. 
 In this paper  we will study the transport coefficients that occur in the parity odd sector, 
 at second order in the derivative expansion.   These are terms constructed out of various
 derivatives in the fluid variables which are odd under parity. 
 We will  consider relativistic fluid systems  with  one additional conserved current. 

 It is usually  difficult to compute  
transport coefficients from the microscopic theory and they are generically determined from experiments.
However parity odd transport coefficients which occur at the first order in the derivative 
 expansion  have been 
 related to quantum anomalies of the microscopic theory \cite{Son:2009tf,Banerjee:2008th,Erdmenger:2008rm,
Loganayagam:2012pz,Jensen:2012jy,Banerjee:2012cr,Loganayagam:2012zg,Sadofyev:2010is} 
\footnote{See \cite{Landsteiner:2012kd} for a recent review  with a complete
list of references.}. 
 Our goal is  to see if a similar phenomenon occurs at the second order in the derivative
 expansion of the constitutive relations.  One  motivation to consider 
 second order fluid dynamics is that  first order fluid dynamics 
 is known to have problems with causality and numerical stability. 
 Transport coefficients which occur at second order provide important constraints 
 for spectral densities  through  sum rules \cite{Romatschke:2009ng}. 
Parity odd transport coefficients   
 at second order affect the dispersion relation of chiral  modes \cite{PhysRevLett.104.198301}
\footnote{We thank Yashodhan Hatwalne for bringing this reference to our attention.}. 
 This phenomenon has important experimental consequences like the
spatial  separation of particles of different chirality.  
 In the relativistic context this phenomenon was first observed holographically 
\cite{Sahoo:2009yq} and then 
 understood due to the presence of a parity odd transport coefficient 
at second order by \cite{Kharzeev:2011ds}.

 Parity odd transport coefficients at second order has been studied earlier in \cite{Kharzeev:2011ds}
 for conformal fluids. 
They used  the principle that parity odd terms which are even under time-reversal 
invariance should not contribute to local entropy production. 
With this principle they  could  constrain these transport 
coefficients and determine some of them. 
We will use the method developed in \cite{Banerjee:2012iz, Jensen:2012jh} to determine and constrain the 
parity odd transport coefficients. We  consider non-conformal fluids in $3+1$ dimensions 
which admits one anomalous  charge current. 
This method  is based  on the requirement that the fluid equations have to be consistent 
with the existence of an equilibrium partition function. Therefore the approach first relies on the 
physical requirement of the existence of equilibrium. 
More precisely: 
\begin{itemize}
\item
 In a time independent background, 
 that is  a space-time metric with a time like Killing vector 
and background gauge fields independent of the time direction, 
any fluid equation will admit a time independent solution. 
\end{itemize}
 The second assumption is:
\begin{itemize}
 \item 
  The stress tensor and the charge current evaluated on this 
 time independent solution can be obtained from the partition function by varying it 
 with respect to the background metric and the gauge field. 
\end{itemize}
This method is implemented as follows. 
\begin{enumerate}
\item
We first  classify all the parity odd 
transport coefficients till  the second order in the derivative expansion of the 
stress tensor and the charge current using symmetries. 
\item
We then  evaluate the stress tensor and the charge current
 on the equilibrium fluid configuration 
to the second order in the derivative expansion. 
\item
The equilibrium partition function is written to the  second order in  derivative expansion
taking all the parity odd terms into consideration.  This is also done based on symmetries
\item 
This stress tensor  and the charge 
current obtained from the equilibrium partition function is 
required to agree with that obtained from the stress tensor evaluated on the 
equilibrium fluid configuration. 
\end{enumerate}

From the description of the  method it is clear that only transport coefficients which do not vanish 
in the equilibrium fluid configuration will be constrained or determined.
We will see that in total there are 27 parity odd transport coefficients. Out of these
12 do not vanish in the equilibrium fluid configuration. 
Among the 12, we determine 7 which we label as $\Phi_i, i=1, \cdots 6$ and $\Delta_2$. 
 and show that they are related to the anomaly. 
The rest of the $5$ are constrained by 3 relations. These relations  also involve  the 
anomaly. 
The results are summarized in (\ref{concludsum}) and (\ref{relationsum}). 
We will then  derive  Kubo formulae for two  of the transport coefficients $\Phi_1, \Phi_2$
and show that it agrees with the equilibrium partition function method. 
The remaining transport coefficients seem to be related to three point functions.

The organization of the paper is as follows. 
In the rest of the introduction we summarize our main results. 
In section 2 we implement the method of \cite{Banerjee:2012iz} to relate the transport coefficients which do not 
vanish in equilibrium to  the anomalies. 
In section 3 we use the Kubo formalism to derive two of the transport coefficients. 
In section 4 we study the effects of the second order  transport coefficients on 
linearized dispersion relations. We also 
 verify the relation between the transport coefficient $\Phi_1$ and the anomaly 
coefficient obtained using the holographic evaluation of this transport coefficient for the 
case of ${\cal N}=4$ Yang-Mills.
Using holography we also show that for the conformal case of ${\cal  N}=4$ Yang-Mills 
the transport coefficient $\Phi_{12}$ vanishes.  In appendix A we discuss 
some of the details involving the classification of the parity odd data at 
second order in derivatives. In appendix B we show the consistency of 
the velocity profile used to derive the Kubo formulae for transport coefficients.

\subsection {Summary of the results}\label{summary}

As we mentioned earlier, the aim of this note is to  constrain the parity odd second order transport coefficients of
an anomalous charged fluid in the presence of  background 
electric and magnetic fields.
To define the transport coefficients unambiguously we first must have an 
 unambiguous definition of fluid variables, that is the velocities, the temperature and the chemical
potentials. 
 $\{u^\mu, T, \mu\}$.
 We will  work in Landau frame which is defined by the following two conditions for
the  charged fluid.
\begin{equation}
J^\mu u_\mu = -q,\qquad T^{\mu\nu}u_\mu = -E u^\nu.
\end{equation}
Let us now consider the  expansion of the charge current 
 $J^\mu$ and  the stress tensor $T^{\mu\nu}$ in terms of the number of 
space time derivatives. This is given by 
\begin{equation}
J^\mu = J_{(0)}^\mu + J_{(1)}^\mu + \cdots, \qquad T^{\mu\nu} = T_{(0)}^{\mu\nu} +T_{(1)}^{\mu\nu} +\cdots,
\end{equation}
where the subscript $(i)$  refer to the number of space-time derivatives. 
The terms $J_{(i)}^\mu $ and $T_{(i)}^{\mu\nu}$ 
for $i\neq 0$ are all perpendicular to the velocity $u^\mu$.
The equations of motion for the fluid 
in the  presence of external electromagnetic field  are given by the  
following conservation laws
\begin{eqnarray}\label{eom1}
\nabla_\mu T^{\mu\nu} &=& {\cal  F}^{\nu\mu}J_\mu
+ \frac{c_m}{2}\nabla_\mu[ \epsilon^{\alpha\beta\gamma\delta} {\cal  F}_{\alpha\beta} R^{\mu\nu}_{\;\gamma\delta}],
\\ \nonumber
 \nabla_\mu J^\mu &=& -\frac{C}{8}\epsilon^{\alpha\beta\gamma\delta}{\cal  F}_{\alpha\beta} {\cal  F}_{\gamma\delta} + 
\frac{c_m}{4} \epsilon^{\alpha\beta\gamma\delta} R^\mu_{\nu\alpha\beta}R^\nu_{\mu\gamma\delta} \\ \nonumber &=& C E_\mu B^\mu 
+ \frac{c_m}{4} \epsilon^{\alpha\beta\gamma\delta} R^\mu_{\nu\alpha\beta}R^\nu_{\mu\gamma\delta}.
\end{eqnarray}
Here $E_\mu = {\cal  F}_{\mu\nu}u^\nu$,  $B^\mu = \frac{1}{2}\epsilon^{\mu\nu\alpha\beta}u_\nu {\cal  F}_{\alpha\beta}$ and $C$ is
the gauge anomaly coefficient and $c_m$ is the coefficient of the 
mixed  gauge-gravitational anomaly.  Note that the terms proportional to the mixed anomaly 
are fourth order in derivatives, therefore they do not affect the analysis of the equations of 
motion to 2nd order in derivatives.  However the gravitational anomaly does enter the 
discussion of the equilibrium partition function at the first order in the derivative 
expansion \cite{Jensen:2012kj}. 

We will now state the known results for the form of the stress tensor and the current
up to first order in the derivative expansion. 
For $i = 0$, the part  with no space-time derivative the stress tensor and current is completely determined by thermodynamics.
At  first order in derivative expansion, that is 
$i=1$, the form of the current and stress tensor
is explicitly known \cite{Banerjee:2012iz}, \cite{Son:2009tf}. 
This form  is consistent with all physical requirements   in the  presence of anomaly  
as well as an external electromagnetic field \cite{Banerjee:2012iz}, \cite{Son:2009tf}. 
The final result for the stress tensor and the charge current to first order in 
derivatives is given by 
\begin{equation}\label{prothom}
\begin{split}
[T_{(0)}]_{\mu\nu} =&~ (E + P) u_\mu u_\nu + P G_{\mu\nu},\\
[T_{(1)}]_{\mu\nu} =&~ -2\eta \sigma^{\mu\nu} - \zeta \Theta P^{\mu\nu},\\
J_{(0)}^\mu =&~ q u^\mu,\\
J_{(1)}^\mu =& ~\Delta V^\mu +\xi_l~ l^\mu + \xi_B B^\mu.\\
\end{split}
\end{equation}
Here $ E$ is the  energy density, $P$ the  pressure and $q$, the  charge density  and 
$G_{\mu\nu}$ is the background metric. 
The variables 
$\Theta, ~\sigma^{\mu\nu},~V^\mu,~l^\mu$ and $ B^\mu$ are all on-shell independent
terms which are  first order in derivatives. These are defined in 
table \ref{table:1storder}. 
The variables 
$\eta,~\zeta,~\Delta,~\xi_l$ and $\xi_B$ refer to the first order transport coefficients.

\begin{table}[ht]
\vspace{0.5cm}
\centering % used for centering table
\begin{tabular}{|c| c| c| c|} % centered columns (4 columns)
\hline\hline %inserts double horizontal lines
Scalars &Vectors & Pseudo Vectors & Tensors \\
(1) & (3) & (2) & (1)\\ [1ex] % inserts table
%heading
\hline % inserts single horizontal line
\hline
$\Theta =\nabla_\mu u^\mu$ & $u^\mu\nabla_\mu u_\nu$ & $l^\mu = \epsilon^{\mu\nu\alpha\beta}u_\nu\partial_\alpha u_\beta$ & $\sigma_{\mu\nu} =\nabla_{\langle\mu} u_{\nu\rangle}$ \\
& $P^{\mu\nu}\nabla_\nu \nu$ & $B^\mu = \frac{1}{2}\epsilon^{\mu\nu\alpha\beta}u_\nu {\cal  F}_{\alpha\beta}$ &\\
& $V^\mu = \left(\frac{E^\mu}{T} - P^{\mu\nu}\nabla_\nu \nu\right)$ &&\\
\hline
\hline
\end{tabular}\vspace{.5cm}
\caption{Data at 1st order in derivative}
\label{table:1storder} % is used to refer this table in the text
\end{table}
\noindent
Throughout this paper,  the symbol $A_{\langle\mu\nu\rangle}$ 
on any tensor $A_{\mu\nu}$ 
denotes the projected, traceless, symmetric part of the tensor.
\begin{equation}
A_{\langle\mu\nu\rangle}= P_\mu^\alpha P_\nu^\beta \left(\frac{A_{\alpha \beta} + A_{\beta\alpha}}{2} - 
\frac{P^{\gamma\theta}A_{\gamma\theta}}{3}G_{\alpha\beta}\right),
\end{equation}
and $P^{\mu\nu}$ is the projector
\begin{equation}
 P^{\mu\nu} = u^\mu u^\nu + G^{\mu\nu}.
\end{equation}
The variable $\nu$ is related to the  chemical potential  by 
\begin{equation}
 \nu = \frac{\mu}{T}.
\end{equation}

Let us now proceed to the stress tensor and the charge current 
at second order in derivatives which we denote as  $[T_{(2)}]_{\mu\nu}$ and $J_{(2)}^\mu$. 
Purely from symmetry considerations, 
 the number of independent transport coefficients upto second order is equal to
total number of possible scalars which appear in the trace of the stress tensor, 
together with the number of possible vectors which appear in the 
current and the possible symmetric traceless tensors which appear 
in the traceless part of the stress tensor. 
But not all of them are independent, they can be related using the equations of motion 
given in (\ref{eom1}). 
In Table \ref{table:2ndorder} we have listed all the parity odd and on-shell independent scalars, 
vectors and tensors containing two space-time derivatives. Some of the details that went into
this classification is discussed in appendix A.

\begin{table}[ht]
 % title of Table
\vspace{0.5cm}
\centering % used for centering table
\begin{tabular}{|c| c| c| c|} % centered columns (4 columns)
\hline\hline %inserts double horizontal lines
Pseudo-scalars  & Pseudo Vectors & Pseudo-tensors \\
(6) & (9) & (12) \\ [1ex] % inserts table
%heading
\hline % inserts single horizontal line
\hline
${\cal S}_1=l^\mu\partial_\mu \nu$ & ${\cal V}_{(1)}^\mu=\epsilon^{\mu\nu\alpha\beta}u_\nu B_\alpha l_\beta$ & $\tau^{(1)}_{\mu\nu}=\nabla_{\langle\mu} l_{\nu\rangle}$\\
${\cal S}_2=B^\mu\partial_\mu \nu$ & ${\cal V}_{(2)}^\mu=\epsilon^{\mu\nu\alpha\beta}u_\nu (\partial_\alpha \nu)(\partial_\alpha T)$ & $\tau^{(2)}_{\mu\nu}= \nabla_{\langle\mu} B_{\nu\rangle}$\\
${\cal S}_3=l^\mu\partial_\mu T$& ${\cal V}_{(3)}^\mu=\Theta l^\mu$ & $\tau_{\mu\nu}^{(3)}=l_{\langle\mu} \partial_{\nu\rangle}\nu$\\
${\cal S}_4=B^\mu\partial_\mu T$& ${\cal V}_{(4)}^\mu=\epsilon^{\mu\nu\lambda\sigma}u_\nu \nabla_\lambda V_\sigma$ & $\tau_{\mu\nu}^{(4)}=B_{\langle\mu} \partial_{\nu\rangle}\nu$ \\
${\cal S}_5=l^\mu V_\mu $& ${\cal V}_{(5)}^\mu=\sigma^\mu_\nu l^\nu$  & $\tau_{\mu\nu}^{(5)}=l_{\langle\mu} \partial_{\nu\rangle}T$\\
${\cal S}_6=B^\mu V_\mu $& ${\cal V}_{(6)}^\mu=\Theta B^\mu$ &$\tau_{\mu\nu}^{(6)}=B_{\langle\mu} \partial_{\nu\rangle}T$ \\
& ${\cal V}_{(7)}^\mu=\sigma^\mu_\nu B^\nu$  & $\tau_{\mu\nu}^{(7)}=l_{\langle\mu}V_{\nu\rangle}$\\
& ${\cal V}_{(8)}^\mu=\epsilon^{\mu\nu\alpha\beta}u_\nu (\partial_\alpha T) V_\beta$ & $\tau_{\mu\nu}^{(8)}=B_{\langle\mu}V_{\nu\rangle}$\\
& ${\cal V}_{(9)}^\mu=\epsilon^{\mu\nu\alpha\beta}u_\nu (\partial_\alpha \nu) V_\beta$ & $\tau_{\mu\nu}^{(9)}=u_\theta \sigma_{\langle\mu\alpha}(\partial_\beta T){\epsilon^{\theta \alpha\beta}}_{\nu\rangle}$\\
&  & $\tau_{\mu\nu}^{(10)}=u_\theta \sigma_{\langle\mu\alpha}(\partial_\beta \nu){\epsilon^{\theta \alpha\beta}}_{\nu\rangle}$\\
& &$\tau_{\mu\nu}^{(11)}=u_\theta \sigma_{\langle\mu\alpha}V_\beta\epsilon^{\theta \alpha\beta}_{\nu\rangle}$\\
& &$\tau_{\mu\nu}^{(12)}=u_\theta \nabla_\beta\sigma_{\langle\mu\alpha}{\epsilon^{\theta \alpha\beta}}_{\nu\rangle}$\\
\hline
\hline
\end{tabular}\vspace{.5cm}
\caption{Parity odd data at 2nd order in derivatives}
\label{table:2ndorder} % is used to refer this table in the text
\end{table}

From the table it can be seen  that  at second order in derivatives there are
 27 parity odd transport coefficients which  appear  in  the current and the stress tensor. 
Therefore the  most general parity odd contributions at 
second order in the  stress tensor and current can be parametrized as follows.
\begin{equation}\label{dwitio}
\begin{split}
[T_{(2)}]_{\mu\nu} =&~\sum_{i = 1}^{12} \Phi_i~ \tau^{(i)}_{\mu\nu}
+P_{\mu\nu} \left[ \sum_{i = 1}^6\chi_i~ {\cal S}_i\right],\\
J_{(2)}^\mu =& ~\sum_{i = 1}^{9}\Delta_i~{\cal V}_{(i)}^\mu.
\end{split}
\end{equation}
Where $\tau^{(i)}_{\mu\nu},~{\cal S}_i$ and ${\cal V}_{(i)}^\mu$ are defined in Table \ref{table:2ndorder}. 
Our goal is to constrain the transport coefficients 
$\Phi_i,~\chi_i$ and $\Delta_i$ ,
using the existence of an equilibrium partition function. 
It can be shown among these 27 terms only 12 can be non-zero in a time independent 
equilibrium fluid configuration. 
These are the first $4$  in the list of scalars ${\cal S}_1\cdots {\cal S}_4$, 
the first $2$ in the list of vectors ${\cal V}^\mu_{(1)}, {\cal V}^\mu_{(2)}$ and 
the first $6$ in the list of tensors $\tau_{\mu\nu}^{(1)} \cdots \tau_{\mu\nu}^{(6)}$. 
Therefore the analysis using the equilibrium partition function can be used to 
constrain the $12$ transport coefficients multiplying these non-vanishing terms. 
These are $\chi_1\cdots \chi_4$ and $\Delta_1, \Delta_2$ and $\Phi_1 \cdots \Phi_6$. 
The final result of this analysis is the following. 
\begin{equation}\label{concludsum}
\begin{split}
&\Phi_1 =  \eta ~b_1,~~\Phi_2 = 2 \eta ~b_2,~~
\Phi_3 = \eta\left(\frac{\partial b_1}{\partial\nu}\right),
~~\Phi_4 =2\eta\left(\frac{\partial b_2}{\partial\nu}\right),\\
&\Phi_5=\eta\left[-\frac{b_1}{T}+\frac{\partial b_1}{\partial T}\right],~~~
\Phi_6 =2\eta\left[-\frac{b_2}{T}+\frac{\partial b_2}{\partial T}\right],
\end{split}
\end{equation}
\begin{equation}\label{relationsum}
\begin{split}
&\Delta_2 =-\frac{\Delta b_1}{2},\\
&T^2 R_1\left[\chi_3- \frac{\zeta}{2}\left(\frac{\partial b_1}{\partial T}- \frac{2 b_1}{T}\right)\right] -R_2\left[\chi_1 - \frac{\zeta}{2}\left(\frac{\partial b_1}{\partial\nu} - 2 b_2 T\right)\right]=0,\\
&T^2 R_1\left[\chi_4- \zeta\left(\frac{\partial b_2}{\partial T}- \frac{b_2}{T}\right)\right] +R_2\left[\chi_2 - \zeta\left(\frac{\partial b_2}{\partial\nu} \right)\right]=0,\\
&R_1 T \Delta_1 + \left[\chi_2 - \zeta\left(\frac{\partial b_2}{\partial\nu} \right)\right] - \frac{q }{ (E + P)}\left[\chi_1 - \frac{\zeta}{2}\left(\frac{\partial b_1}{\partial \nu} - 2 b_2 T\right)\right] = 0,
\end{split}
\end{equation}
where
$$b_1 = \frac{T^3}{E + P} \left(\frac{2C\nu^3}{3}- 4 C_2 \nu\right)
,~~b_2 =  \frac{T^2}{E + P} \left(\frac{C\nu^2}{2}-C_2\right),$$
$$R_1 = \left(\frac{\partial P}{\partial E}\right)_q,~~R_2 = \left(\frac{\partial P}{\partial q}\right)_E.$$
Note that $C$ is the  gauge anomaly coefficient and  $C_2$ is related to the coefficient of the 
mixed gauge-gravitational anomaly\cite{Jensen:2012kj} by \footnote{The relation \eqref{mgrav} is derived in \cite{Jensen:2012kj} using properties of the partition function on cones. 
It is an equation that relates coefficients at different orders in derivative expansion. In our analysis we shall simply assume their result. We will subsequently see that this 
identification is also consistent with the analysis of \cite{Landsteiner:2012kd} which 
studies  the effect of gravitational anomalies on hydrodynamics.}
\begin{equation} \label{mgrav}
C_2 = 8\pi^2 c_m.
\end{equation}
 Therefore  we see that the coefficients
$\Phi_1, \cdots \Phi_6$ and $\Delta_2$ are  determined in terms of the  anomaly, the shear viscosity $\eta$,
the charge diffusivity $\Delta$ and the thermodynamic functions $E, P, T,  \nu$. 
The rest $\Delta_1, \chi_1, \cdots \chi_4$ are constrained by $3$ relations which 
involve the anomaly and the bulk viscosity $\zeta$.

\section{Anomalous transport from equilibrium partition function }\label{implement}

In  sub-section \ref{genmethod} we will  briefly outline the general procedure 
we use  to relate  parity odd transport  at the second order in derivatives
to the anomaly. 
This method has been described  and analyzed in detail by 
\cite{Banerjee:2012iz} and \cite{Jensen:2012jh}.  It relies on the analysis of the 
equilibrium partition function. Thus we will be able to constrain only those transport 
coefficients which do not vanish in the equilibrium configuration. 
This section also will serve to introduce the notation and conventions used in the paper. 
In  sub-section \ref{appl} we implement this method and derive the 
relations given in (\ref{concludsum}) and (\ref{relationsum}). 

\subsection{The equilibrium partition function method}\label{genmethod}

 We are interested in a fluid flow on a static background metric and
  a static external electromagnetic field. 
  The most general static metric and gauge field can be written in the following form.
 \begin{equation}\label{back}
\begin{split}
ds^2 &= G_{\mu\nu} dx^\mu dx^\nu = -e^{2\sigma}(dt + a_i dx^i)^2 + g_{ij} dx^i dx^j,\\
\text{Gauge Field}:~~& {\cal A}_\mu dx^\mu,
\end{split}
\end{equation}
$\sigma$, $a_i$ , $g_{ij}$, ${\cal A}_0$ and ${\cal A}_\mu$ are all slowly varying functions of 
the spatial co-ordinates $(\vec x)$.
Our notations are as follows:
\begin{itemize}
\item Greek indices run from 1 to 4.
\item Latin indices run from 1 to 3.
\item All Greek indices are lowered or raised by the 4 dimensional metric $G_{\mu\nu}$ unless explicitly mentioned.
\item All Latin indices are lowered or raised by the 3 dimensional metric $g_{ij}$ unless explicitly mentioned.
\item $\bar\nabla_\mu$ is covariant derivative with respect to the metric $G_{\mu\nu}$ and $\nabla_i$ is the covariant derivative with respect to the metric $g_{ij}$. 
\item For any tensor $A_{\mu\nu}$ the notation $A_{\langle\mu\nu\rangle}$ denotes the traceless symmetric part of the tensor, projected in direction perpendicular to the fluid velocity.
\begin{equation*}
A_{\langle\mu\nu\rangle}= P_\mu^\alpha P_\nu^\beta \left(\frac{A_{\alpha \beta} + A_{\beta\alpha}}{2} - \frac{P^{\gamma\theta}A_{\gamma\theta}}{3}G_{\alpha\beta}\right),
\end{equation*}
where $P_{\mu\nu} \equiv u_\mu u_\nu + G_{\mu\nu}$ is referred to as the projector. 
\end{itemize}

Our basic assumption is that
in such a background any fluid equation will admit a time independent solution. The stress tensor and the current, evaluated on this time independent solution, can be generated by varying the partition function of the system with respect to the background metric and gauge field.
 If $Z$ is the partition function at temperature $T_0$  then the stress tensor and the current evaluated on the  equilibrium  are given by the following formulae.
\begin{equation}\label{formula}
\begin{split}
&T_{00}|_{equilibrium} = -\frac{T_0 e^{2\sigma}}{\sqrt{-G}}\left[\frac{\delta Z}{\delta \sigma}\right] = \frac{T_0^2}{\sqrt{g}}\left[\frac{\delta Z}{\delta \bar T}\right],\\
&T^i_0|_{equilibrium} =  \frac{T_0 }{\sqrt{-G}}\left[\frac{\delta Z}{\delta a_i} - A_0 \frac{\delta Z}{\delta A_i}\right] = \frac{\bar T }{\sqrt{g}}\left[\frac{\delta Z}{\delta a_i} - A_0 \frac{\delta Z}{\delta A_i}\right],\\
&T^{ij}|_{equilibrium}= -\frac{2T_0 }{\sqrt{-G}} g^{il}g^{jm}\left[\frac{\delta Z}{\delta g^{lm}  }\right]= -\frac{2\bar T }{\sqrt{g}}g^{il}g^{jm}\left[\frac{\delta Z}{\delta g^{lm}  }\right],\\
&J_0|_{equilibrium} = -\frac{T_0 e^{2\sigma}}{\sqrt{-G}}\left[\frac{\delta Z}{\delta A_0}\right] = -\frac{ e^{\sigma}}{\sqrt{g}}\left[\frac{\delta Z}{\delta \bar\nu}\right],\\
&J^i|_{equilibrium} = \frac{T_0 }{\sqrt{-G}} \left[\frac{\delta Z}{\delta A_i  }\right] = \frac{\bar T }{\sqrt{g}} \left[\frac{\delta Z}{\delta A_i  }\right],
\end{split}
\end{equation}
where $$\bar T = T_0 e^{-\sigma},~~\bar \nu = \frac{A_0}{T_0},~~A_i = {\cal A}_i - A_0 a_i.$$

The strategy we will adopt to constrain the parity odd coefficients which occur at the 
second order is the following:
\begin{enumerate}
\item Write down the most general partition function upto a given order in derivative expansion and consistent with all the symmetries.
It will be a function of $a_i$, $\sigma$ and ${\cal A}_i$ and  their derivatives.

\item Vary the partition function  $Z$,  to obtain the most general possible expression for $T^{\mu\nu}|_{\rm equilibrium}$ 
and $J^\mu|_{\rm equilibrium}$.

\item Parametrize the most general possible fluid stress tensor and current up to some given order in derivative expansion using symmetries.  
This will give the maximum number of independent transport coefficients possible constrained only by symmetry. 
We have stated the results of this analysis already  in equations \eqref{prothom} and \eqref{dwitio}.

\item Evaluate the most general fluid stress tensor and current on the equilibrium solution. 
The final outcome will contain some of the unknown transport coefficients.

 \item Equate the final outcome of the previous step with the stress tensor and current we have already  obtained by varying the partition function.
 
\item This will express the transport coefficients  which appear in the  stress tensor and the current evaluated in equilibrium, in terms of the 
coefficients appearing in the partition function.

\item Eliminating the coefficients which appear in the partition function results in  the desired relations among the transport coefficients. 

\end{enumerate}

One might wonder that to execute the fourth step,  the  equilibrium solution for the velocity, temperature and  the other fluid variables needs 
to be independently found. 
But  as it has been explained in \cite{Banerjee:2012iz}  using this method one can perturbatively determine  both the solution and the transport
coefficients in terms of the free functions appearing in the partition function. 

\subsection{Partition function analysis: parity-odd sector at second order}\label{appl}

In this subsection we shall apply the general procedure described in \ref{genmethod} to the particular case of the  parity-odd sector of the 
charged fluid at second order in derivative expansion.
The first contribution to the parity-odd sector comes at first order in derivative expansion. This has been analyzed in detail in  section 3 of
\cite{Banerjee:2012iz}. We will
 not repeat the first order computation here but we will extensively use their result.

 \subsubsection*{Stress tensor and current from the  partition function}
Since the transport coefficients we are interested in belong to the parity-odd sector, 
we  will restrict our attention to only the parity odd part of the partition function.
The  partition function $Z_{(2)}$ at second order in 
derivatives is a gauge invariant scalar functional of the background metric and gauge-fields.\footnote{ Though the system is anomalous,
all the effects of anomaly i.e. the anomalous transformation property of the partition function under the gauge transformation
is  accounted by the  first order part $Z_{(1)}$. Therefore  in $Z_{(2)}$ we need to consider only  gauge invariant scalars.}
Hence we need to list   all possible parity odd scalars that can be constructed from the metric functions and gauge fields that contain 
two space-derivatives.
Note that since all the functions are time independent no  time derivatives occur. 
There are four such scalars:
\begin{enumerate}
\item $\epsilon^{ijk}\partial_i {\bar\nu} f_{jk}$,
\item $\epsilon^{ijk}\partial_i {\bar T} f_{jk}$,
\item $\epsilon^{ijk}\partial_i {\bar\nu} F_{jk}$,
\item $\epsilon^{ijk}\partial_i {\bar T} F_{jk}$,
\end{enumerate}
where $\bar \nu = \frac{A_0}{T_0},~~\bar T = T_0 e^{-\sigma}, ~~A_i = {\cal A}_i - A_0 a_i$
and $f_{jk} = \partial_j a_k - \partial_k a_j,~~F_{jk} = \partial_j A_k - \partial_k A_j$. 

Therefore naively, the  parity odd second order partition function 
at two derivative order can have 4 free parameters, but two of them can be related by total derivatives. Ignoring the total 
derivative terms the most general second order partition function can be written as
$$Z_{(2)} = \int \sqrt{g} \left[ M_1 (\bar T, \bar \nu)~\epsilon^{ijk}\partial_i {\bar\nu} F_{jk} + T_0M_2 (\bar T,\bar\nu)~ \epsilon^{ijk}\partial_i {\bar\nu} f_{jk}\right].$$
Varying this partition function with respect to the metric and the gauge fields and using \eqref{formula} 
we get the following second order correction to the stress tensor and charge current.
\begin{equation}\label{strscr}
\begin{split}
[\Pi^{(2)}]_{00} &= T_0^2 \left[\left(\frac{\partial M_1}{\partial \bar T}\right)\epsilon^{ijk}\partial_i {\bar\nu} F_{jk} + T_0 \left(\frac{\partial M_2}{\partial \bar T}\right) \epsilon^{ijk}\partial_i {\bar\nu} f_{jk}\right],\\
[\Pi^{(2)}]^i_0 &= 2 T_0 \bar T \left( \frac{\partial M_2}{\partial \bar T} - \bar \nu \frac{\partial M_1}{\partial \bar T}\right) \epsilon^{ijk}(\partial_j\bar T)(\partial _k \bar \nu),\\
[\Pi^{(2)}]^{ij} &= 0,\\
\\
[j^{(2)}]_0 &= \frac{T_0}{\bar T} \left[\left(\frac{\partial M_1}{\partial \bar T}\right)\epsilon^{ijk}\partial_i {\bar T} F_{jk} + T_0\left( \frac{\partial M_2}{\partial \bar T}\right)\epsilon^{ijk}\partial_i {\bar T} f_{jk}\right],\\
[j^{(2)}]^i &=2 {\bar T}\left(\frac{\partial M_1}{\partial \bar T}\right)\epsilon^{ijk}(\partial_j\bar T)(\partial _k \bar \nu).
\end{split}
\end{equation} 
It is  important  to note the third equation in 
\eqref{strscr}. The fact that the  spatial components of the equilibrium stress tensor 
vanish at second order will serve as an important constraint in determining the 
transport coefficients. 

\subsubsection*{Stress tensor from fluid dynamics}

We have to evaluate the fluid dynamical stress tensor on the equilibrium, that is a time independent solution in the given static
background metric and gauge field. This equilibrium solution for the velocity field, temperature or chemical potential in terms of the 
background metric and gauge field  can also be expanded in terms of derivatives. We shall use the following notation.
\begin{equation}\label{notation3}
\begin{split}
u^\mu|_{eq} &= \bar u^\mu + \delta u^\mu_{(1)} + \delta u^\mu_{(2)} + \cdots,\\
T|_{eq} &= \bar T + \delta T_{(1)} + \delta T_{(2)} + \cdots,\\
\nu|_{eq} &=  \bar\nu + \delta\nu _{(1)} + \delta \nu_{(2)} + \cdots,
\end{split}
\end{equation}
where
$\delta u^\mu_{(i)}$, $\delta T_{(i)}$ and $\delta\nu _{(i)}$ are $i$th corrections to the zeroth order equilibrium solution containing 
$i$ derivatives on the background data.
Now we have to substitute \eqref{notation3} in fluid stress tensor and current given in \eqref{prothom}) and extract the part that will 
be parity odd and  which  involve exactly two derivatives on the background data. This is the part which have to be equated with \eqref{strscr}.

From the analysis done in \cite{Banerjee:2012iz} we know that\footnote{Note that equation \eqref{correction} is valid only if we choose Landau frame. 
This is the place where the choice of a fluid frame enters our analysis. We are going to use these equations in all our subsequent calculation.}
\begin{equation}\label{correction}
\begin{split}
&\bar u^\mu = e^{-\sigma} \{1,0,0,0\}, ~~\bar T = T_0e^{-\sigma} ,~~\bar \nu = \frac{A_0}{T_0},\\
&[\delta u_{(1)}]_0 =0,~~\delta T_{(1)} =0,~~\delta\nu_{(1)} = 0,\\
& [\delta u_{(1)}]^0 = - a_i [\delta u_{(1)}]^i ,\\
&[\delta u_{(1)}]^i = \left(\frac{b_1}{2}\right) \bar l^i + b_2 \bar B^i,\\
&[\delta u_{(1)}]_i = g_{ij} [\delta u_{(1)}]^j,
\end{split}
\end{equation}
where 
\begin{equation}\label{note2}
\begin{split}
&F_{jk}  \equiv \partial_j A_k - \partial_k A_j,\\
&\bar l^i = -\frac{e^{\sigma}}{2}\epsilon^{ijk}f_{jk},\\
&\bar B^i = \frac{1}{2}\left(\epsilon^{ijk}F_{jk} +A_0\epsilon^{ijk}f_{jk}\right) =\frac{1}{2}\epsilon^{ijk}F_{jk} -\bar T\bar\nu\bar l^i , \\
&b_1 = \frac{T^3}{E + P} \left(\frac{2C\nu^3}{3}- 4 C_2 \nu\right),\\
&b_2 =  \frac{T^2}{E + P} \left(\frac{C\nu^2}{2}-C_2\right).
\end{split}
\end{equation}
Here $C$ is the anomaly coefficient and $C_2$ is related to the mixed
anomaly\footnote{The relation \eqref{rr} was derived in \cite{Jensen:2012kj}.  Here we simply
use it. } in (\ref{eom1}) by
\begin{equation}\label{rr}
C_2 = 8\pi^2 c_m.
\end{equation}

If we also expand $E$, $P$ and $q$  in terms of a derivative expansion as
$$E|_{eq} = \bar E + \delta E_{(1)} + \delta E_{(2)} + \cdots,~~P|_{eq} = \bar P + \delta P_{(1)} + \delta P_{(2)} + \cdots,~~q|_{eq} = \bar q + \delta q_{(1)} + \delta q_{(2)} + \cdots,\\$$
then from \eqref{correction} it follows that
\begin{equation}\label{c2}
\delta E_{(1)} = \delta P_{(1)} = \delta q_{(1)}=0.
\end{equation}
Using the fact that $(u_\mu u^\mu = -1)$ to all order in derivative expansion we find that 
\begin{equation}\label{c3}
[\delta u_{(2)}]_0 \propto  [\delta u_{(1)}]^i [\delta u_{(1)}]_i = \text{Parity even}\sim 0~~ (\text{For our purpose}) .
\end{equation} 
Also using the Landau gauge condition on the
 second order stress tensor and current one can show that in equilibrium
\begin{equation}\label{c4}
[T_{(2)}]_{00} = [T_{(2)}]_0^i = [J_{(2)}]_0=0.
\end{equation}
Note that $[T_{(2)}]_{00}, [T_{(2)}]_0^i , [J_{(2)}]_0$ are the components of the 
stress tensor and the charge current which are second order in derivatives obtained
from the equation (\ref{dwitio}). While $[\Pi^{(2)}]_{00},  [\Pi^{(2)}]_0^i,  [j^{(2)}]_0$ 
refer to the components of the stress tensor and the charge current
obtained from the equilibrium partition function using (\ref{strscr}). \\
Using \eqref{correction}, \eqref{c2}, \eqref{c3} and \eqref{c4} we get the following form for the second order stress
tensor and current evaluated in equilibrium.
\begin{equation}\label{duideri}
\begin{split}
[\Pi^{(2)}]_{00}&=\delta E _{(2)} \bar u_0^2  + \delta(-2\eta \sigma_{00}) + \delta(-\zeta \Theta P_{00}),\\
[\Pi^{(2)}]_0^i&= (\bar E + \bar P) \bar u_0[\delta u_{(2)}]^i + \delta(-2\eta \sigma_0^i) + \delta(-\zeta \Theta P_0^i),\\
[j^{(2)}]_0 &= \delta q_{(2)} \bar u_0 + \delta(\Delta V_0)+\delta (\xi_l l_0) + \delta (\xi _B B_0),
\end{split}
\end{equation}
where $\delta(-2\eta\sigma_{\mu\nu})$, $\delta(\zeta \Theta P_{\mu\nu})$  and $\delta (\Delta V_\mu)$ are the two derivative corrections of $(-2\eta\sigma_{\mu\nu})$,
$(\zeta \Theta P_{\mu\nu})$ and $(\Delta V_\mu)$ when evaluated on $\delta u^\mu_{(1)}$, $\delta T_{(1)}$ and $\delta \nu _{(1)}$.

Now if $Q^{(1)}_{\mu\nu}$ is a tensor 
which is first order in the derivative expansion satisfying the following two conditions
\begin{equation}
Q^{(1)}_{\mu\nu}u^\mu = 0~~ \text{ at all orders and} \qquad Q^{(1)}_{\mu\nu}|_{(\bar u^\mu, \bar T,\bar\nu)} = 0,
\end{equation}
then  one can show
 in general that  
 \footnote{$0=\delta \left[u^\mu Q^{(1)}_{\mu\nu}\right]=\bar u^\mu \delta Q^{(1)}_{\mu\nu}+\delta u^\mu \left[Q^{(1)}_{\mu\nu}|_{(\bar u^\alpha, \bar T,\bar\nu) }\right]= e^{-\sigma}\delta Q^{(1)}_{0\nu}$.}
\begin{equation} \delta Q^{(1)}_{00} = \delta [Q^{(1)}]_0^i = 0.
\end{equation}
Similarly if $Q^{(1)}_\mu$  is a vector which is  first order in the  derivative expansion satisfying 
\begin{equation}
Q^{(1)}_\mu u^\mu = 0~~ \text{ at all orders and } \qquad
Q^{(1)}_\mu|_{(\bar u^\mu, \bar T,\bar\nu)} = 0,
\end{equation}
then  it follows that
\footnote{$0=\delta [u^\mu Q^{(1)}_\mu]=\bar u^\mu \delta Q^{(1)}_\mu+\delta u^\mu [Q^{(1)}_\mu|_{(\bar u^\alpha, \bar T,\bar\nu) }]= e^{-\sigma}\delta Q^{(1)}_0$.}
\begin{equation}
\delta Q^{(1)}_0 = 0.
\end{equation}
 This  argument  allows us to conclude that 
$$\delta(-2\eta \sigma_{00}) =\delta(\zeta \Theta P_{00}) = \delta(-2\eta \sigma_0^i) = \delta(\zeta \Theta P_0^i)=\delta(\Delta V_0)=0.$$
Similarly  $\delta (\xi_l l_\mu)$ and  $\delta(\xi _B B_\mu)$ are the two derivative corrections of $ (\xi_l l_\mu)$ and 
$(\xi _B B_\mu)$ when evaluated on $\delta u^\mu_{(1)}$, $\delta T_{(1)}$ and $\delta \nu _{(1)}$. But $l_\mu$ and $B_\mu$ are already parity odd. 
Therefore $\delta (\xi_l l_\mu)$ and $\delta (\xi_B B_\mu)$
are going to be parity even and hence we can set them to zero for our purpose.
Putting all this together we have the following result for the components of the stress tensor and the
current at the second order in derivative expansion
\begin{equation}\label{duifinal}
\begin{split}
[\Pi^{(2)}]_{00}&=\delta E _{(2)} \bar u_0^2 = e^{2\sigma}\delta E _{(2)},\\
[\Pi^{(2)}]_0^i&= (\bar E + \bar P)\bar u_0 [\delta u_{(2)}]^i = -e^\sigma (\bar E + \bar P) [\delta u_{(2)}]^i,\\
[j^{(2)}]_0 &= -e^\sigma\delta q_{(2)}. \\
\end{split}
\end{equation}
Inverting \eqref{duifinal} and using \eqref{strscr} we can determine the second order piece of the equilibrium solution for velocity, 
temperature and chemical potential in terms of background data.
Now the components $[\Pi^{(2)}]^{ij}$ and $[j^{(2)}]^i$ will  give rise to  the constraints on the transport  coefficients.
\begin{equation}\label{tranfin}
\begin{split}
[\Pi^{(2)}]^{ij} &= \delta P_{2} g^{ij} -2\eta~\delta\sigma^{ij} - \zeta~\delta \Theta g^{ij} + [T_{(2)}]^{ij},\\
[j^{(2)}]^i &= \bar q [\delta u_{(2)}]^i +\Delta~\delta[ V^i]+[J_{(2)}]^i . 
\end{split}
\end{equation}
We have to use \eqref{duifinal} to determine $\delta P_2$ and $[\delta u_{(2)}]^i$.

The transport coefficients are determined by demanding  that 
\eqref{tranfin} is satisfied. For convenience let us further split the first equation in \eqref{tranfin} in two parts, the trace part
which is obtained by 
 contracting the equation with $g_{ij}$ and a traceless part
 which is obtained by  subtracting the trace part of the  equation from the first equation of  \eqref{tranfin}
 \footnote{Note that here trace is taken with the 3 dimensional metric.}.
 These are given by   
\begin{equation}\label{trtrl}
\begin{split}
&{\rm Trace~part:}\\ &\delta P_{2} - \zeta\delta \Theta + 
\frac{1}{3}(-2\eta\delta\sigma^{ij} + [T_{(2)}]^{ij})g_{ij}=0.\\
\\
&{\rm Traceless~part:} \\& -2\eta\left(\delta\sigma^{ij} -\frac{g^{ij}}{3}( \delta\sigma^{lm}g_{lm})\right) +\left([T_{(2)}]^{ij} -\frac{g^{ij}}{3}( [T_{(2)}]^{lm}g_{lm})\right)=0.
\end{split}
\end{equation}
In \eqref{trtrl} we have used \eqref{strscr} to set $[\Pi^{(2)}]^{ij}$ to zero.  Therefore the RHS of \eqref{trtrl} vanishes.

\subsubsection*{Transport  in the traceless part of the stress tensor}
In this subsection we shall analyze the second equation of \eqref{trtrl}.
\begin{equation}\label{sip}
\begin{split}
&{\rm Traceless~part:}\\& -2\eta\left(\delta\sigma^{ij} -\frac{g^{ij}}{3}( \delta\sigma^{lm}g_{lm})\right) +\left([T_{(2)}]^{ij} -\frac{g^{ij}}{3}( [T_{(2)}]^{lm}g_{lm})\right)=0.
\end{split}
\end{equation}
We now go through the following steps:
\begin{itemize}
\item We  first evaluate  the second order contribution from $\sigma^{ij}$.
\item  We then write down the most general form for  $[T_{(2)}]^{\langle\mu\nu\rangle}$  from usual symmetry analysis and on shell independence. In general it will contain many unknown transport transport coefficients to begin with. But if we restrict our attention to  only  the parity odd ones, the total number of terms is  18 ( $\Phi_i$, $(i = 1,\cdots,12)$ multiplying 12 independent traceless symmetric tensors and $\chi$, $(i = 1,\cdots,6)$ multiplying 6 independent scalars, see \eqref{dwitio}). All these 18 terms have been listed in table \ref{table:2ndorder}. 
\item One can see that all the 6 $\chi_i$s are not going to appear in the combination that we are going to evaluate in \eqref{sip}.  
\item So we have to evaluate the rest of the 12 terms (multiplying $\Phi_i$s on the equilibrium time-independent solution. 

For this it is sufficient to substitute only the zeroth order equilibrium solution, since each of these terms already contain two derivatives.
\item As mentioned in section 1.1, it turns out that 6 of these twelve terms evaluate to zero in equilibrium. So there are only 6 transport coefficients which can be constrained by this equilibrium analysis.
These 6 terms are the following.
\begin{eqnarray}
\tau^{(1)}_{\mu\nu}=\bar\nabla_{\langle\mu }l_{\nu\rangle},~~\tau^{(2)}_{\mu\nu}=\bar\nabla_{\langle\mu} B_{\nu\rangle},~~\tau^{(3)}_{\mu\nu}=(\partial_{\langle\mu}\nu) l_{\nu\rangle}, \\ \nonumber
\tau^{(4)}_{\mu\nu}=(\partial_{\langle\mu}\nu) B_{\nu\rangle},~~\tau^{(5)}_{\mu\nu}=(\partial_{\langle \mu}T) l_{\nu\rangle},~~\tau^{(6)}_{\mu\nu}=(\partial_{\langle \mu}T) B_{\nu\rangle},
\end{eqnarray}
where $l^\mu \equiv \epsilon^{\mu\nu\alpha\beta}u_\nu \partial_\alpha u_\beta$ and $ B^\mu \equiv \epsilon^{\mu\nu\alpha\beta}u_\nu \partial_\alpha {\cal A}_\beta$. 
Let us now  rewrite the traceless part of the second order fluid stress tensor with these six terms, this is 
given by
\begin{equation}
[T_{(2)}]_{odd}^{\mu\nu} = \sum_{a = 1}^6\Phi_a[\tau^{(a)}]^{\mu\nu} + \text{ trace part} + \text{terms that vanish in equilibrium}.
\end{equation}
\item  From \eqref{sip} we see that each 
of these six $\Phi_a$'s has to be such that, they cancel the contribution of  
$\delta \sigma_{\mu\nu}$ when all of them are evaluated in equilibrium.
\item Evaluating the spatial components of these six terms on the zeroth order equilibrium 
solution  given in (\ref{correction}) we obtain
\begin{equation}\label{evaluate}
\begin{split}
&[\tau^{(1)}]^{ij}=\bar\nabla^{\langle i}l^{j\rangle} = g^{il} g^{jm}\bigg[\frac{\nabla_l \bar l_m + \nabla_m \bar l_l}{2} - \frac{g_{lm}}{3}(\nabla_k\bar l^k)\bigg] + {\cal O}(\partial^3),\\
&[\tau^{(2)}]^{ij}=\bar\nabla^{\langle i}B^{j\rangle} = g^{il} g^{jm}\bigg[\frac{\nabla_l \bar B_m + \nabla_m \bar B_l}{2} - \frac{g_{lm}}{3}(\nabla_k\bar B^k)\bigg]+ {\cal O}(\partial^3),\\
&[\tau^{(3)}]^{ij}=(\bar\nabla^{\langle\mu}\nu) l^{\nu\rangle}= g^{il} g^{jm}\bigg[\frac{(\nabla_l \bar\nu) \bar l_m + (\nabla_m \bar \nu)\bar l_l}{2} - \frac{g_{lm}}{3}(\nabla_k\bar\nu\bar l^k)\bigg]+ {\cal O}(\partial^3),\\
&[\tau^{(4)}]^{ij}=(\bar\nabla^{\langle\mu}\nu) B^{\nu\rangle}= g^{il} g^{jm}\bigg[\frac{(\nabla_l \bar\nu) \bar B_m + (\nabla_m \bar \nu)\bar B_l}{2} - \frac{g_{lm}} {3}(B^k\nabla_k\bar\nu\bar )\bigg]+ {\cal O}(\partial^3),\\
&[\tau^{(5)}]^{ij}=(\bar\nabla^{\langle \mu}T) l^{\nu\rangle}= g^{il} g^{jm}\bigg[\frac{(\nabla_l \bar T) \bar l_m + (\nabla_m \bar T)\bar l_l}{2} - \frac{g_{lm}}{3}(\bar l^k\nabla_k\bar T)\bigg]+ {\cal O}(\partial^3),\\
&[\tau^{(6)}]^{ij}=(\bar\nabla^{\langle \mu}T) B^{\nu\rangle}= g^{il} g^{jm}\bigg[\frac{(\nabla_l \bar T) \bar B_m + (\nabla_m \bar T)\bar B_l}{2} - \frac{g_{lm}}{3}(\bar B^k\nabla_k \bar T)\bigg]+ {\cal O}(\partial^3),\\
\end{split}
\end{equation}
where $\bar l^i \equiv -\frac{e^\sigma}{2}\epsilon^{ijk}f_{jk}$ and  $\bar B^i \equiv \frac{1}{2}\epsilon^{ijk}(F_{jk} + A_0 f_{jk})$. To evaluate equation \eqref{evaluate} we have 
extensively used section 2 of  \cite{Banerjee:2012iz}.
\item The spatial  components  of the shear tensor gives the following contribution to $[\Pi^{(2)}]^{ij}$
\begin{equation}\label{shearcont}
\begin{split}
&-2 \eta ~\delta\sigma^{ij}\\
=&-2\eta\bigg[\bar P^{i\mu} \bar P^{j\nu}\left\{\frac{\bar\nabla_\mu [\delta u_{(1)}]_\nu + \bar\nabla_\nu [\delta u_{(1)}]_\mu}{2} - \frac{\bar\nabla_\alpha[\delta u_{(1)}]^\alpha}{3}\right\}\\
&~~~~~~~~~~~~+\left(\bar P^{j\mu}\delta P^{i\nu} + \bar P^{i\nu}\delta P^{j\mu}\right)\left(\frac{\bar\nabla_\mu \bar u_\nu + \bar\nabla_\nu \bar u_\mu}{2}\right)\bigg]\\
 =& -2\eta \bigg[g^{il} g^{jm}\left\{\frac{\nabla_l [\delta u_{(1)}]_m + \nabla_m [\delta u_{(1)}]_l}{2} - \frac{g_{lm}}{3}\nabla_k[\delta u_{(1)}]^k- \frac{g_{lm}}{3}(\partial_k\sigma)[\delta u_{(1)}]^k\right\}\\
         &~~~~~~~~~~~~~~~ +\frac{\delta u^i g^{jk}(\partial_k\sigma) + [\delta u_{(1)}]^j g^{ik}(\partial_k\sigma)}{2}\bigg],\\
\end{split}
\end{equation}
where 
$$\bar P^{i\mu} = \bar u^i \bar u^\mu + G^{i\mu} = G^{i\mu},~~
\bar P^{j\mu}\delta P^{i\nu} = G^{j\mu}\bar u^\nu[\delta u_{(1)}]^i.$$
The last term in the last line of \eqref{shearcont} results  from this expansion of the projectors.
Substituting \eqref{correction} in \eqref{shearcont} we obtain the following
\begin{equation}\label{shearcontex}
\begin{split}
-2 \eta \delta\sigma^{ij} &= -2\eta\bigg[ \frac{b_1}{2}[\tau^{(1)}]^{ij} +b_2[\tau^{(2)}]^{ij}
+\frac{1}{2}\left(\frac{\partial b_1}{\partial\nu}\right)[\tau^{(3)}]^{ij}  +\left(\frac{\partial b_2}{\partial\nu}\right)[\tau^{(4)}]^{ij}\\
&~~~~~~~~~~~ +\frac{1}{2}\left(-\frac{b_1}{T}+\frac{\partial b_1}{\partial T}\right)[\tau^{(5)}]^{ij}
+\left(-\frac{b_2}{T}+\frac{\partial b_2}{\partial T}\right)[\tau^{(6)}]^{ij}\bigg].
\end{split}
\end{equation}
\end{itemize}

\noindent
Next we have to substitute \eqref{shearcont} and \eqref{shearcontex} in \eqref{sip}. Now to satisfy \eqref{sip} the coefficient of each independent expression should vanish.  From examining 
\eqref{shearcont} and \eqref{shearcontex} it seems that demanding every independent 
term to vanish results in more equations than the unknowns which are the 
transport coefficients. However from the 
structure of the equations in \eqref{sip} it is clear that the equation 
admits the following unique solution for the transport coefficients
\begin{equation}\label{conclud}
\begin{split}
&\Phi_1 =  \eta ~b_1,~~\Phi_2 = 2 \eta ~b_2,~~
\Phi_3 = \eta\left(\frac{\partial b_1}{\partial\nu}\right),
~~\Phi_4 =2\eta\left(\frac{\partial b_2}{\partial\nu}\right),\\
&\Phi_5=\eta\left[-\frac{b_1}{T}+\frac{\partial b_1}{\partial T}\right],~~~
\Phi_6 =2\eta\left[-\frac{b_2}{T}+\frac{\partial b_2}{\partial T}\right],
\end{split}
\end{equation}
where 
\begin{equation}
b_1 = \frac{T^3}{E + P} \left(\frac{2C\nu^3}{3}- 4 C_2 \nu\right),
~~b_2 =  \frac{T^2}{E + P} \left(\frac{C\nu^2}{2}-C_2\right),
\end{equation}
and $E, P, T$ is the energy density, pressure and temperature respectively. 
$\nu = \frac{\mu}{T}$ refers to the chemical potential, and $\eta$ is the shear viscosity. 
$C$ is the gauge anomaly coefficient, while $C_2$ is related  to the mixed 
anomaly by (\ref{mgrav}).  The transport coefficient $\Phi_1$ gives rise to chiral 
dispersion relations in the shear mode \cite{Kharzeev:2011ds}.

\subsubsection*{Transport in the trace of the stress tensor and current}
We will now constrain the transport coefficients that occur in the trace part of the stress
tensor and the current using the equilibrium partition function. 
We will first rewrite the relevant part of the current and the trace part of the stress tensor up to 2nd order in derivative expansion. 
Using \eqref{strscr} and \eqref{tranfin}   we obtain
the following equation for the trace part of the stress tensor and current at second order.
\begin{equation}\label{sami2}
\begin{split}
{\rm Trace~part:}~~ 
0&=\delta P_{2} - \zeta\delta \Theta +\frac{1}{3}(-2\eta\delta\sigma^{ij} + [T_{(2)}]^{ij})g_{ij}\\
&=\delta P_{2} - \zeta\delta \Theta + \left[\sum_{1}^6\chi_i{\cal S}_i\right]_{eq}\\
&=\left [-\zeta \delta\Theta + \left(\frac{\partial P}{\partial E}\right)_q \delta E_{(2)} +\left(\frac{\partial P}{\partial q}\right)_E \delta q_{(2)}\right]+\left[\sum_{1}^6\chi_i{\cal S}_i\right]_{eq}\\
\end{split}
\end{equation}
\begin{equation}\label{sami3}
\begin{split}
{\rm Current:}~~
[J_{(2)}]^i
 &= [j^{(2)}]^i-\delta [\Delta V^i] - q [\delta u_{(2)}]^i\\
\Rightarrow \left[\sum_{k =1}^{9}\Delta_k {\cal V}^i_{(k)}\right]_{eq}&= 2\bar T \left(\frac{\partial M_1}{\partial \bar T}\right)\epsilon^{ijk}(\partial_j\bar T)(\partial_k\bar\nu)-\delta [\Delta V^i] - q [\delta u_{(2)}]^i.
\end{split}
\end{equation}
Both in \eqref{sami2} and \eqref{sami3} we have used \eqref{dwitio} to write the fluid  stress tensor and the current  in terms of the transport coefficients.
Also in \eqref{sami2} to evaluate the combination $(-2\eta\delta\sigma^{ij} + [T_{(2)}]^{ij})g_{ij}$  we have used \eqref{shearcontex} and \eqref{conclud}. One can see that the transport coefficients $\Phi_i$s drop out. 

Using \eqref{duifinal} and \eqref{strscr} we evaluate 
$\delta E_2$, $\delta q_2$ and $[\delta u_{(2)}]^i$, this results in 
\begin{equation}\label{equ}
\begin{split}
&\delta E_2 =e^{-2\sigma}\Pi^{(2)}_{00} = 2 \bar T^2 \left(\frac{\partial M_1}{\partial \bar T}\right)(\bar B^i \partial_i\bar\nu) - 2 \bar T^3\left(\frac{\partial M_2}{\partial\bar T} - \bar \nu \frac{\partial M_1}{\partial\bar T}\right)(\bar l^i \partial_i\bar\nu),\\
&\delta q_2 = e^{-\sigma}j_0^{(2)}= 2 \left(\frac{\partial M_1}{\partial \bar T}\right)(\bar B^i\partial_i\bar T) - 2\bar T \left(\frac{\partial M_2}{\partial\bar T} - \bar\nu \frac{\partial M_1}{\partial\bar T}\right) (\bar l^i\partial_i\bar T),\\
&[\delta u_{(2)}]^i = -\left(\frac{ e^{-\sigma}}{\bar E + \bar P}\right)[\Pi^{(2)}]^i_0 = - \left(\frac{2\bar T^2}{E + P}\right)\left(\frac{\partial M_2}{\partial\bar T} - \bar \nu \frac{\partial M_1}{\partial\bar T}\right)\epsilon^{ijk}(\partial_j\bar T)(\partial _k \bar \nu).
\end{split}
\end{equation}
Now we have to compute $\delta\Theta$ and $\delta(\Delta V^i)$.
\begin{equation}\label{dtheta}
\begin{split}
\delta\Theta =~& \bar\nabla_\mu [\delta u_{(1)}]^\mu =\nabla_k [\delta u_{(1)}]^k - \left(\frac{\partial_k \bar T}{\bar T}\right)[\delta u_{(1)}]^k \\
=~&\frac{1}{2}\left(\frac{\partial b_1}{\partial \bar T} - \frac{2 b_1}{\bar T}\right) (\bar l^i\partial_i \bar T) + \frac{1}{2}\left(\frac{\partial b_1}{\partial\bar\nu} - 2 b_2 \bar T\right)(\bar l^i\partial_i \bar \nu)\\
&+\left(\frac{\partial b_2}{\partial \bar T} - \frac{ b_2}{\bar T}\right) (\bar B^i\partial_i \bar T) + \left(\frac{\partial b_2}{\partial\bar\nu} \right)(\bar B^i\partial_i \bar \nu).\\
\end{split}
\end{equation}
To derive \eqref{dtheta} we have used the following identities
\begin{equation}
\nabla_i\bar l^i = -\left(\frac{\partial_i\bar T}{\bar T}\right)\bar l^i,~~~~\nabla_i\bar B^i = -\bar T\bar l^i\partial_i\bar \nu. 
\end{equation}
Similarly $\delta V^i$ is given by the following expression
\begin{equation}\label{dV}
\begin{split}
&\delta\{\Delta V^i\} = \Delta G^{i\mu}{\cal F}_{\mu\nu}\delta u^\nu =\Delta\epsilon^{ijk} [\delta u_{(1)}]_j\bar B_k = \Delta\left(\frac{b_1}{2}\right)\epsilon^{ijk}\bar l_j \bar B_k, \\
\end{split}
\end{equation}
where ${\cal F}_{\mu\nu} =\partial_\mu {\cal A}_\nu - \partial_\nu{\cal A}_\mu$. 

Now we have to evaluate the 6 scalars (${\cal S}_i,~~i = 1, \cdots, 6$) and the $i$th component of the 9 vectors (${\cal V}_{(k)}^i,~~k = 1,\cdots,9$) on the equilibrium solution. We need  the equilibrium solution only at zeroth order since all of them  already contain two derivatives. Explicit expressions for all these terms are listed in table \ref{table:2ndorder}. 

As mentioned in section 1.1, only 6 terms, 4 scalars and 2 vectors are 
non-vanishing on the zeroth order equilibrium solution. 
Thus the  relevant parts of the second order current and trace of the stress tensor 
are given by 
\begin{equation}\label{repeat}
\begin{split}
[T_{(2)}]^{\alpha\beta}|_{trace~ part}&=P^{\alpha\beta}\left[ \chi_1 (l^\mu\partial_\mu \nu) + \chi_2 (B^\mu\partial_\mu \nu) +\chi_3( l^\mu\partial_\mu T) + \chi_4 (B^\mu\partial_\mu T)\right],\\
J_{(2)}^\mu &= ~\Delta_1 [\epsilon^{\mu\nu\alpha\beta}u_\nu (\partial_\alpha \nu)(\partial_\beta T)]+ \Delta_2 [ \epsilon^{\mu\nu\alpha\beta}u_\nu B_\alpha l_\beta] 
.\end{split}
\end{equation}
We then evaluate all these six terms on the zeroth order equilibrium solution.
\begin{equation}\label{evtrace}
\begin{split}
&(l^\mu\partial_\mu \nu)= (\bar l^i\partial_i\bar \nu),\\
&(B^\mu\partial_\mu \nu)= (\bar B^i\partial_i \bar\nu),\\
&(l^\mu\partial_\mu T)= (\bar l^i\partial_i\bar T),\\
&(B^\mu\partial_\mu T)= (\bar B^i\partial_i \bar T),\\
&\epsilon^{i\nu\alpha\beta}u_\nu (\partial_\alpha \nu)(\partial_\beta T)=\epsilon^{ijk}(\partial_j\bar \nu) (\partial_k\bar T),\\
& \epsilon^{i\nu\alpha\beta}u_\nu B_\alpha l_\beta = \epsilon^{ijk} \bar B_j \bar l_k.
\end{split}
\end{equation}

Substituting \eqref{evtrace} in \eqref{repeat} we  express the LHS of \eqref{repeat} in terms of the background data and the two arbitrary functions $M_1$ and $M_2$ of the second order parity odd partition function. Now using \eqref{sami2} and \eqref{strscr} we can express the 5 transport coeffcients appearing in \eqref{repeat} in terms of $M_1$ and $M_2$. Since one cannot generate a term of the form $\epsilon^{ijk}l_j B_k$ from the partition function, the 6th transport coefficients 
$\Delta_2$ will be completely determined by the correction of the first order current.
This is similar to the  way the transport coefficients $\Phi_i$'s,  which
appear  in the traceless part of the stress tensor  were determined.
The end result of this step is
\begin{equation}\label{m1m2}
\begin{split}
&\Delta_2 = -\frac{\Delta ~b_1}{2},\\ 
&\chi_1 = -2 R_1 \bar{ T}^3 \left[\bar\nu\frac{\partial M_1}{\partial \bar T} -\frac{\partial M_2}{\partial\bar T}\right] + \zeta\left(\frac{1}{2}\frac{\partial b_1}{\partial\nu} - b_2 \bar T\right) ,\\
&\chi_2 = -2R_1 \bar{T}^2\left(\frac{\partial M_1}{\partial \bar T}\right)+\zeta\left(\frac{\partial b_2}{\partial\nu}\right),\\
&\chi_3 = -2R_2\bar{ T} \left[\bar\nu\frac{\partial M_1}{\partial \bar T} -\frac{\partial M_2}{\partial \bar T}\right] +\frac{\zeta}{2}\left(\frac{\partial b_1}{\partial\bar T}- \frac{2b_1}{\bar T}\right), \\
&\chi_4 = 2R_2 \left(\frac{\partial M_1}{\partial \bar T}\right) +\zeta\left(\frac{\partial b_2}{\partial \bar T} - \frac{b_2}{\bar T}\right),\\
&\Delta_1 = 2\bar T\left(\frac{\partial M_1}{\partial \bar T}\right) + \left(\frac{2 q T^2}{E + P}\right)\left(\frac{\partial M_2}{\partial \bar T}- \bar\nu \frac{\partial M_1}{\partial \bar T}\right), 
\end{split}
\end{equation}
where
 $$R_1 = \left(\frac{\partial P}{\partial E}\right)_q,~~R_2 = \left(\frac{\partial P}{\partial q}\right)_E,$$
$$b_1 = \frac{T^3}{E + P} \left(\frac{2C\nu^3}{3}- 4 C_2 \nu\right)
,~~b_2 =  \frac{T^2}{E + P} \left(\frac{C\nu^2}{2}-C_2\right).$$
Eliminating $M_1$ and $M_2$ from these expressions we get three relations among the remaining 
5  second order transport coefficients.
\begin{equation}\label{relation2}
\begin{split}
&\Delta_2 =-\frac{\Delta b_1}{2},\\
&T^2 R_1\left[\chi_3- \frac{\zeta}{2}\left(\frac{\partial b_1}{\partial T}- \frac{2 b_1}{T}\right)\right] -R_2\left[\chi_1 - \frac{\zeta}{2}\left(\frac{\partial b_1}{\partial\nu} - 2 b_2 T\right)\right]=0,\\
&T^2 R_1\left[\chi_4- \zeta\left(\frac{\partial b_2}{\partial T}- \frac{b_2}{T}\right)\right] +R_2\left[\chi_2 - \zeta\left(\frac{\partial b_2}{\partial\nu} \right)\right]=0,\\
&R_1 T \Delta_1 + \left[\chi_2 - \zeta\left(\frac{\partial b_2}{\partial\nu} \right)\right] - \frac{q }{ (E + P)}\left[\chi_1 - \frac{\zeta}{2}\left(\frac{\partial b_1}{\partial \nu} - 2 b_2 T\right)\right] = 0.
\end{split}
\end{equation}

\section{Kubo formula for the transport coefficients $\Phi_1, \Phi_2$}
\label{Kubo}

In this section we derive the  
relations obtained for the transport coefficients 
$\Phi_1$ and $\Phi_2$ given in  (\ref{conclud}) using the Kubo formula.
We consider the following equilibrium  background for the metric, gauge field and the 
velocity 
\begin{equation} \label{bcke}
g_{\mu\nu}^{(0)} = {\rm diag} ( -1, 1, 1, 1), \quad
{\cal A}_{\mu} = ( \nu^{(0)} T_0, 0 , 0 , 0) , \quad
u^{\mu} = ( 1, 0, 0, 0) .
\end{equation}
The chemical potential $\nu^{(0)}$ and the temperature $T_0$  are constants and do not depend on space-time.  Since the energy $E^{(0)}$ and the pressure $P^{(0)}$ can be thought of  as functions 
of  the temperature and the chemical potential, they are also constants in space-time. 
Now consider  the following  non-zero metric perturbations  about this background 
\begin{equation}
\delta g_{tx} = h_{tx}, \quad \delta g_{tz} = h_{tz}, \quad \delta g_{yx} = h_{yx}, \quad \delta 
g_{yz} = h_{yz}.
\end{equation}
The gauge field perturbations are given by 
\begin{equation}
\delta {\cal A}_\mu = (0, a_x, 0, a_z).
\end{equation}
The  fluid velocity is close to the rest frame and its perturbations are given by 
\begin{equation} \label{velpert}
\delta u_\mu = ( 0, v^x,  0, v^z) .
\end{equation}
All perturbations are assumed to have dependence only in the time $t$ and $y$-direction. 
In  appendix \ref{console} we will show that the background and the perturbations considered in 
equations (\ref{bcke}) to (\ref{velpert}) consistently solve the linearized fluid equations without the 
need for turning on any other perturbations. A simple reason that these 
perturbations consistently solve the linearized fluid equations is that they 
are all in the spin-2 shear sector and therefore they  decouple from the rest. 

To derive Kubo formulae for transport coefficients, 
we consider the constitutive relations for the stress tensor and the 
charge current as one point functions in the presence of external sources. 
We then obtain two point functions for the currents by differentiating with respect to the
metric and the gauge field perturbations. 
Working this out to the linear order in perturbations will result in Kubo formulae for the 
transport coefficients $\Phi_1, \Phi_2$. 

To proceed we will require the the Christoffel symbols to the linear order in perturbations. 
The non-vanishing elements at the linear order are given by 
\begin{eqnarray}
\label{eqn3.5}
\Gamma^t_{xy} = -\frac{1}{2} \left( \partial_y h_{tx} - \partial_t h_{yx} \right) , \qquad
\Gamma^{t}_{zy} = - \frac{1}{2} \left( \partial_y h_{tz} -\partial_t h_{zy} \right), 
\\ \nonumber
\Gamma^x_{tt}  = \partial_t h_{tx} , \quad 
\Gamma^x_{ty} = \frac{1}{2} \left( \partial_y h_{tx} + \partial_t h_{yx} \right), 
\quad  \Gamma^x_{yy} = \partial_y h_{xy}, \\ \nonumber
\Gamma^z_{tt} = \partial_t h_{z0}, \quad 
\Gamma^z_{ty} = \frac{1}{2} \left( \partial_y h_{tz} + \partial_t h_{yz} \right), 
\quad  \Gamma^x_{yy} = \partial_y h_{zy}, \\ \nonumber
\Gamma^y_{tx} = \frac{1}{2}  \left( \partial_t h_{yx} - \partial_y h_{tx} \right), 
\qquad
\Gamma^y_{tz} = \frac{1}{2} \left( \partial_t  h_{yz} - \partial_y h_{tz} \right).
\end{eqnarray}
Evaluating the inverse metric to the linear order  we obtain 
\begin{equation}
h^{tx} = h_{tx}, \quad h^{tz} = h_{tz}, \quad h^{yx} = - h_{yx}, \quad h^{yz} = - h_{yz}.
\end{equation}
The covariant components of the velocity  are given by 
\begin{equation}
u_{\mu} = ( -1, v^x + h_{xt} , 0, v^z + h_{zt} ) .
\end{equation}

We now evaluate  various components of the stress tensor  to the linear order in 
the perturbations. 
From the list of terms that contribute at 2nd order in derivatives given in table 2 
we see that in the background we have chosen all the scalars ${\cal S}_i$ vanish. 
The reason for this is for the background all the thermodynamic functions are 
independent of space-time. 
We now examine  the traceless part of the stress tensor. Note that the contributions from 
$\tau_{\mu\nu}^{(i)}$ for $i = 3, 4, 5, 6, 7, 8, 9, 10, 11$ vanish 
since the thermodynamic functions are independent of space-time. 
What remains to be evaluated are the contributions from $\tau_{\mu}^{(i)} $ for
$i=1, 2$ and $i=12$.  
Let us first examine the $tx$ and the $ty$ component of the stress tensor. 
To the second order in derivatives and to the linear order in the perturbation this is given by 
 \begin{eqnarray}\label{stress1}
 T^{tx} &=& ( E^{(0)}  + P^{(0} ) v^x + P^{(0)}  h_{tx} ,  \\ \nonumber
T^{tz} &=& (E^{(0)} + P^{(0)}  ) v^z + P^{(0)}  h_{tz} , \\ \nonumber
T^{ty} &=& 0 .
\end{eqnarray}
Note that $\sigma^{tx}, \sigma^{tz}$, 
$[\tau^{(i)}]^{tx}$ and $[\tau^{(i)}]^{tz}$  for $i=1, 2, 12$ do not contribute at the linear order. 
The reason is that the term $ \nabla _{ \alpha} u_{\beta}$ 
and $\nabla_{ \alpha } l_{\beta} $ is a first order term, 
therefore one has to evaluate the projector for these components say 
$P^{t\alpha}P^{x\beta}$ at the zeroth order, which  vanishes. 

Lets examine the $yx$  component of the stress tensor. 
We need to evaluate the contributions from $[\tau^{(i)}]^{yx}$ for $i =1, 2, 12$. 
These are given by  \footnote{In evaluating these contributions we take $\epsilon^{0123} = \frac{1}{\sqrt{-g} } $. } 
\begin{eqnarray}
 [\tau^{(1)}]^{yx} &=&  \frac{1}{2} \partial_y^2 ( v^z + h_{zt}), \\ \nonumber
  [\tau^{(2)}]^{yx} &=& \frac{1}{2} \partial_y^2 a_z, \\ \nonumber
[\tau^{(12)}]^{yx} &=& \frac{1}{2} \partial_y (  \partial_ y v^z  + \partial_t h_{yz} ) .
\end{eqnarray}
We also need the contribution of the shear tensor to the linear order. 
This is given by 
\begin{equation}
 \sigma^{yx} = \frac{1}{2} ( \partial_yv^x + \partial_t h_{yx}) .
\end{equation}
Considering all these contributions along with the contribution to the stress tensor to the 
zeroth order in derivative we obtain
\begin{eqnarray}
 T^{yx} &=& -P^{(0)} h_{yx} - \eta( \partial_y v^x + \partial_t h_{yx}) \\ \nonumber
&& +\frac{1}{2}\Phi_1\partial_y^2 ( v^z + h_{zt} ) + \frac{1}{2} \Phi_2 \partial_y^2 a_z 
\\ \nonumber
& & + \frac{\Phi_{12}}{2} \partial_y(  \partial_ y v^z  + \partial_t h_{yz} ).
\end{eqnarray}
The equations of motion for the $x$ component of the stress tensor to the linear order in the 
fields is given by 
\begin{equation}
\partial_tT^{tx} + \partial_t  h_{tx} T^{tt} + \partial_y T^{yx} = 0.
\end{equation}
Here $T^{tt}$ is the zeroth order $tt$ component of the  stress tensor  which is given by 
\begin{equation}
T^{tt} = E^{(0)}.
\end{equation}
Fourier transforming these equations and taking the zero frequency limit or taking the 
time independent situation we obtain 
the Ward identities for the one point function of the stress tensor. 
\begin{equation} \label{wi}
\lim_{\omega\rightarrow 0} T^{yx}(k) = 0.
\end{equation}
We  can now differentiate these with respect to the background fields  $h_{zt}$, and $a_z$
and obtain Kubo formulae for the transport coefficients $\Phi_1$ and $\Phi_2$ respectively. 
A similar procedure for the  $yz$ component of the stress tensor yields the same 
result.

To proceed 
 we first eliminate $v^x$ and $v^z$  using 
(\ref{stress1}).  This results in  the following equations
\begin{eqnarray}
 T^{yx} &=&  -P^{(0)} h_{yx} -\eta\left( \frac{ \partial_y T^{tx} - P^{(0)} 
 \partial_y h_{tx}}{E^{(0)} +P^{(0}}  
 \right)  \\ \nonumber
& & + \frac{\Phi_1}{2} \left( 
\frac{ \partial_y^2 T^{tz} +  E^{(0)}  \partial_y^2 h_{tz} }{E^{(0)} + P^{(0)} } \right)    
 + \frac{\Phi_2}{2} \partial_y^2 a_z \\ \nonumber
& & + \frac{\Phi_{12}}{2}   \left( \frac{ \partial_y^2 T^{tz} - P^{(0)} 
 \partial_y ^2h_{tz}}{E^{(0)} +P^{(0}}  
 \right) .
\end{eqnarray}
Here we have already used time-independence to  drop the time derivatives. 
Fourier transforming these equations we obtain
\begin{eqnarray}
T^{yx}(k) &=& -P^{(0)} h_{yx} - \eta\left( \frac{ik T^{tx}  -ik P^{(0)}  h_{tx} } {E^{(0)} +P^{(0}}
\right)   \\ \nonumber
& & - \frac{\Phi_1}{2} \left( 
\frac{ k^2 T^{tz}  + E^{(0)}  k^2 h_{tz} } {E^{(0)} + P^{(0)} } \right)  
- \frac{\Phi_2}{2} k^2 a_z  \\ \nonumber
& & - \frac{\Phi_{12}}{2} k^2 
\left( \frac{T^{tz} - P^{(0)} h_{tz} }{ E^{(0)} + P^{(0)} }  \right) \\ \nonumber
& =& 0 .
\end{eqnarray}
The last equality in the above equation implements the Ward identity given in (\ref{wi}) and 
$k$ is the momentum in the $y$ direction. 
Let us now focus on the expression for $T^{yx}$ a similar analysis can be repeated for $T^{yz}$. 
Differentiating the Ward identity for $T^{yx}$ with respect to $h_{zt}$ and $a_z$ and setting 
the other backgrounds to zero we obtain  the following two equations
\begin{eqnarray}\label{form1}
& & \frac{ k^2}{E^{(0)} + P^{(0)} }\left[  ( \Phi_1  \langle T^{tz}( k)  T^{tz} ( -k) \rangle  + E^{(0)}  )
+ \Phi_{12} ( \langle T^{tz}( k) T^{tz}(-k) \rangle  -P^{(0)} )  \right] \nonumber\\  
& &\qquad\qquad\qquad\qquad\qquad
 =  - ik \frac{2\eta}{E^{(0)} + P^{(0)} }  \langle T^{tx}(k)  T^{tz}( -k)  \rangle , \\ \nonumber
& & \frac{\Phi_1 + \Phi_{12}}{E^{(0)} + P^{(0)}} k^2  \langle T^{tz}( k) j^z(-k) \rangle + 
 \Phi_2 k^2   
= -ik \frac{2\eta} {E^{(0)} + P^{(0)} } \langle T^{tz} (k) j^{z}(-k) \rangle .
\end{eqnarray}
To obtain the first equation we have differentiated with respect to $h_{tz}$ and set all the other 
backgrounds to zero.  The second equation is obtained by differentiating the Ward identity with 
respect to $a_z$ and setting the remaining backgrounds to zero. 
The equations in (\ref{form1}) are sufficient to determine the transport coefficients 
$\Phi_1, \Phi_2$. 
From \cite{Landsteiner:2012kd} we have the following results for the various
two point functions. 
\begin{eqnarray} \label{2ptfn}
 \lim_{k\rightarrow 0,  \omega\rightarrow 0} \langle T^{tx}(k)  T^{tz}( -k)  \rangle
 &=&  ik \left( \frac{C}{3}(\nu^{(0)}T^{(0)})^3 -  2 C_2    ( T^{(0)})^3 \nu^{(0)}  \right)  , \\ \nonumber
  \lim_{k\rightarrow 0,  \omega\rightarrow 0} \langle T^{tx}(k)  j^{z}( -k)  \rangle
 &=& ik \left(  \frac{C}{2}  (\nu^{(0)} T^{(0)})^2 -  C_2 ( T^{(0)})^2  \right)  .\\ \nonumber
 \end{eqnarray}
 These results are given in equations (79)-(81) for a system with 3 chemical potentials
and equations (123)-(125) for a system with a single chemical potential of 
\cite{Landsteiner:2012kd}. 
The definition of the variables for the two point functions used is given in equation (47). 
This reference also uses the normalization 
\begin{equation}
\frac{-C}{8} = \frac{1}{32\pi^2}, \qquad \frac{c_m}{4}  = \frac{1}{768\pi^2},  \qquad
C_2 = \frac{1}{24},
\end{equation}
for the gauge anomaly and  we have rewritten the chemical potential $\mu$ in terms of the variable
$\nu$. 
From \cite{Amado:2011zx} we can read out the following correlators
 \begin{eqnarray}\label{twopoint}
 \lim_{k \rightarrow 0, \omega\rightarrow  0} \langle T^{tz} (k) T^{tz}(-k) \rangle =  
 P^{(0)}, \qquad
  \lim_{k \rightarrow 0, \omega \rightarrow 0} \langle T^{tz} (k) j^{z} (-k) \rangle  =0.
 \end{eqnarray}
 These correlators are mentioned below equation (2.16) of reference \cite{Amado:2011zx}. 
 Substituting the formulae  for the two point functions given in (\ref{2ptfn}) and (\ref{twopoint}) into
 the equations given in  (\ref{form1}) we obtain 
 \begin{eqnarray}
 \Phi_1 &=& \frac{2\eta}{E^{(0)} + P^{(0)} }\left(  \frac{C}{3}  (\nu^{(0)} T^{(0)} )^3 
 - 2C_2 ( T^{(0)})^3 \nu^{(0)} \right) , \\ \nonumber
 \Phi_2  &=& \frac{2\eta}{E^{(0)} + P^{(0)} } \left( \frac{C}{2}  (\nu^{(0)} T^{(0)} )^2 - C_2 ( T^{(0)})^2  \right). 
 \end{eqnarray}
 These expressions agree with that derived using the equilibrium partition function method 
 which are given in (\ref{conclud}). 
 
 \section{Chiral shear waves}

In this section we examine the effects of the second order parity transport coefficients
on linearized  dispersion relations about the equilibrium characterized by the 
following background given in (\ref{bcke}). 
 Note that the temperature $T_0$ and the 
chemical potential $\nu^{(0)}$ are independent of space-time and therefore all other
thermodynamic variables are constants in space-time. 
We  consider shear modes \footnote{We have shown that none of the 
parity odd transport coefficients modify the sound or the charge dissipation mode for the 
equilibrium characterized by (\ref{bcke}).}, for this we examine 
velocity perturbations of the form
\begin{equation}
\delta u^\mu = ( 0, v^x, 0, v^z) .
\end{equation}
These perturbations depend only on time $t$ and the $y$-direction. 
We include all the terms in the stress tensor to the 2nd order in derivatives 
given by
\begin{equation}
 T^{\mu\nu} = [T_{(0)}]^{\mu\nu} + [T_{(1)}]^{\mu\nu} +  [T_{(2)}]^{\mu\nu}_{odd} \qquad,
\end{equation}
where $[T_{(0)}]^{\mu\nu}$ and  $[T_{(1)}]^{\mu\nu}$ are given in (\ref{prothom}) and 
$[T_{(2)}]^{\mu\nu}_{odd}$ is given in (\ref{dwitio}). 
We now write down the contribution to the stress tensor from these velocity fluctuations which 
are linear order in the fluctuations. 
It can be seen that the only contributions from 2nd order  which arise are from 
the term involving $\Phi_1$ and $\Phi_{12}$.  
Thus the stress tensor to the linear order in velocity fluctuations is given by 
\begin{eqnarray}\label{flutr}
 \delta T^{tx} = ( E^{(0)} + P^{(0)}  ) v^x, \qquad  \delta T^{tz} = ( E^{(0)} + P^{(0)}) v^z, 
\\ \nonumber
\delta T^{yx} =  -\eta\partial_y v^x  + 
\frac{\Phi_1}{2} \partial_y^2 v^z   + \frac{\Phi_{12}}{2} \partial_y^2 v^z, \\ \nonumber
\delta T^{yz} = -\eta\partial_y v^z 
- \frac{\Phi_1}{2} \partial_y^2 v^x   -  \frac{\Phi_{12}}{2} \partial_y^2 v^x.
\end{eqnarray}
We now substitute these values in the equations of motion for the stress tensor given by 
\begin{eqnarray}
\partial_t \delta T^{tx} + \partial_y T^{yx} = 0, \\ \nonumber
\partial_t \delta T^{tz} + \partial_y T^{yz} = 0.
\end{eqnarray}
Substituting the expressions for the stress tensor from (\ref{flutr}) into the above 
equations and then 
taking the Fourier transform of these equations result in the following 
set of coupled equations
\begin{eqnarray}
( -i\omega ( E^{(0)} + P^{(0)} ) + \eta k^2 ) v^x - 
i \frac{k^3}{2} ( \Phi_1 + \Phi_{12})   v^z &=&  0, \\ \nonumber
( -i\omega ( E^{(0)} + P^{(0)} ) + \eta k^2 ) v^z  + 
i \frac{k^3}{2} ( \Phi_1 + \Phi_{12})   v^x &=&  0, \\ \nonumber
\end{eqnarray}
From these equations we see that the two shear modes split depending on their 
chirality. The dispersion relations for these modes 
are given by 
\begin{equation}
 \omega = -i \frac{\eta}{E^{(0)} + P^{(0)} } k^2 
\mp  \frac{i}{2( E^{(0)} +P^{(0)} ) } ( \Phi_1 + \Phi_{12}) k^3 .
\end{equation}
Thus  in the basis 
we have used to list the second order transport coefficients
both  $\Phi_1$ as well as $\Phi_{12}$ contribute to the
splitting of the shear modes. Earlier studies of the chiral shear modes were restricted to the 
conformal transport at second order, therefore the contribution of $\Phi_{12}$ to the splitting 
was not noticed.  Let us call the coefficient of $k^3$ as the chiral dispersion coefficient and 
define
\begin{equation} \label{cdc}
 D = \frac{1}{2( E^{(0)} +P^{(0)} )} ( \Phi_1 + \Phi_{12}) .
\end{equation}

Using holography we now show that for the case ${\cal N}=4$ Yang-Mills,  $\Phi_{12} =0$. 
We will also check the relation of $\Phi_1$ to the anomaly coefficient derived in this 
paper using the holographic result for this transport coefficient. 
The holographic dual of this system  is given by the  Reisner Nordstr\"{o}m black hole 
in $AdS_5$.  We will use the notations of \cite{Banerjee:2008th}. 
The five dimensional action we consider is given by 
\begin{equation}\label{gaction}
S = \frac{1}{16\pi G_5} \int \sqrt{-g_5} \left( R + 12 - F_{AB} F^{AB}
 - \frac{4\kappa}{3} \epsilon^{LABCD} A_L F_{AB} F_{CD} \right) .
\end{equation}
The equations of motion of the above action are
\begin{eqnarray}
G_{AB} - 6 g_{AB} + 2 \left( F_{AC} F^{C}_B + \frac{1}{4} g_{AB} F_{CD} F^{CD} \right) &=&0, 
\\ \nonumber
\nabla_B F^{AB} + \kappa \epsilon^{ABCDE} F_{BC} F_{DE} &=& 0, 
\end{eqnarray}
where $G_{AB}$ is the  five dimensional Einstein tensor. 
The  Reisner-Nordstr\"{o}m black brane solution in Eddington Finkelstein 
coordinates is given by 
\begin{eqnarray} \label{gsol}
ds^2 &=&  -2 u_\mu dx^\mu dr - r^2 V(r, m , q) u_\mu u_\nu dx^\mu dx^\nu + 
r^2 P_{\mu\nu}dx^\mu dx^\nu, \\ \nonumber
A &=& \frac{\sqrt{3} q}{ 2 r^2} u_\mu dx^\mu
\end{eqnarray}
and 
\begin{equation}
 u_\mu dx^\mu = dv , \qquad V(r, m , q) = 1 - \frac{m}{r^4} + \frac{q^2}{r^6}.
\end{equation}
Let $R$ be the radius of the outer horizon of the black hole. We then define the quantities
\begin{equation} \label{scalin}
M \equiv \frac{m}{R^4},  \qquad Q = \frac{q}{R^3},  \qquad Q^2 = M-1.
\end{equation}
The last equation results from the fact that $R$ is the largest root of $V(r) =0$. 
The  thermodynamic quantities of this black hole are given by 
\begin{equation} \label{techem}
T = \frac{R}{2\pi} ( 2- Q^2) , \qquad \mu = 2 \sqrt{3} Q R ,
\end{equation}
where $T$ is the Hawking temperature and $Q$ the charge density. 
The energy density, pressure and the shear  viscosity in terms of these variables are given by 
\begin{equation}\label{ext}
 E^{(0)} = \frac{3 M R^4}{16\pi G_5}, \qquad
P^{(0)} = \frac{M R^4}{ 16\pi G_5}  \qquad
\eta = \frac{ R^3}{ 16\pi G_5} = \frac{s}{4\pi} .
\end{equation}
Before we proceed we first relate the anomaly coefficient $C$ to the Chern-Simons coefficient
$\kappa$. The boundary current is given by 
\begin{equation}
J^\mu = \frac{1}{16\pi G_5} \sqrt{-g_5} F^{r\mu}|_{r\rightarrow\infty},
\end{equation}
where $r$ is the radial direction. Here we are using the definition of the current which 
is consistent with the Page charge using which the charge density of the black hole 
is evaluated. There are other definitions of current as discussed in footnote 3 of
\cite{Banerjee:2008th}.  The bulk equations of motion for the gauge field results in the 
following conservation law for the current. 
\begin{equation}\label{consv}
\partial_\mu J^\mu = - \frac{\kappa}{16\pi G_5} \epsilon^{\mu\nu\rho\sigma} F_{\mu\nu} F_{\rho\sigma}.
\end{equation}
We now have to identify the relation between the  gauge field used in field theory and that of the 
bulk gauge field. Note that the chemical potential value is related to the horizon value of the 
gauge field. Comparing the horizon value of the gauge field in (\ref{gsol}) and (\ref{techem}) we see that 
the relation between the bulk gauge field and the  gauge field is given by 
\begin{equation}\label{scaling}
A_{\mu}^{field} = 4 A_{\mu}^{bulk}.
\end{equation}
The relation is important since we have already used this normalization to define the 
thermodynamics of the boundary gauge theory. The gauge fields  in the field theory 
must be defined consistent with this thermodynamics. 
Substituting the relation (\ref{scaling})  into the conservation law (\ref{consv}) we 
obtain
\begin{equation}
\partial_\mu J^\mu = -\frac{\kappa}{ 256 \pi G_5} \epsilon^{\mu\nu\rho\sigma} F^{field}_{\mu\nu} F^{field}_{\rho\sigma}.
\end{equation}
Now all quantities are defined in the field theory. 
Comparing with the current conservation law in (\ref{eom1}) we obtain
\begin{equation} \label{ancoef}
 C = \frac{\kappa}{32\pi G_5}.
\end{equation}

The transport coefficient $\frac{\Phi_1}{2}$  can be identified to the 
coefficient ${\cal N}_7$ in the notation of \cite{Banerjee:2008th}, equation (4.37)
see also \cite{Megias:2013joa,Erdmenger:2008rm}. Reading out the 
holographic value of  ${\cal N}_7$ from equation (4.38) of \cite{Banerjee:2008th} we have
\begin{equation} \label{holval}
\frac{\Phi_1}{2} = \frac{1}{16\pi G_5} \frac{\sqrt{3}  }{M} ( M-1)^{\frac{3}{2}} R^2 \kappa.
\end{equation}
Note that  the action given in (\ref{gaction})  captures the situation when the 
 gravitational anomaly is  zero.  Using the relations in
(\ref{scalin}),  (\ref{techem}) and (\ref{ext}) it can be seen that $\Phi_1$ can be 
written as 
\begin{equation}
 \Phi_1 =  \frac{\mu^3 \eta}{ E^{(0)} + P^{(0)} } \frac{\kappa}{48 \pi G_5}.
\end{equation}
We can now compare it with  the expression derived earlier for this paper
for $\Phi_1$ for charge fluid with an anomaly which is given by 
\begin{equation}
 \Phi_1 = \frac{2}{3} \frac{\mu^3\eta}{ E^{(0)} + P^{(0)} } C.
\end{equation}
Therefore we obtain
\begin{equation} \label{ancoef2}
 C = \frac{\kappa}{32\pi G_5}.
\end{equation}
This is precisely  the relation between the Chern-Simons coefficient and 
$C$ obtained directly using equations of motion in (\ref{ancoef}). 
This serves as a check for the relation between the transport coefficient $\Phi_1$ 
and the anomaly coefficient derived in this paper. 

We will now use the holographic value of $\Phi_1$ given in (\ref{holval}) to evaluate its contribution 
to the chiral dispersion coefficient for the Riesner-Nordstr\"{o}m black hole 
\begin{equation} \label{hymode}
D =  \frac{\sqrt{3} Q^3 \kappa}{4 M^2 R^2} + \frac{1}{2( E^{(0)} + P^{(0)} )} \Phi_{12}  .
\end{equation}
By the AdS/CFT correspondence the chiral dispersion relation corresponds to the 
quasi-normal modes seen in the shear channel of the graviton fluctuations \cite{Kovtun:2005ev}. 
We therefore  compare this dispersion coefficient with that obtained in \cite{Sahoo:2009yq} by 
studying the  quasi-normal modes in the shear sector. 
They find that the dispersion coefficient is given by 
\begin{equation}\label{qnm}
D^{QNM} = \frac{\kappa^{SY} (Q^{SY})^3 }{8 m^2 R^3}.
\end{equation}
By comparing the action and the background solution  given in \cite{Sahoo:2009yq} to that given in 
(\ref{gaction}) and (\ref{gsol}) we obtain the following relations between the variables 
of \cite{Sahoo:2009yq} and that used here
\begin{equation}
 \kappa^{SY} = \frac{2\kappa}{3},  \qquad
( Q^{SY})^2 = 3 q^2 = 3 Q^2 R^6. 
\end{equation}
Substituting the above relations  in (\ref{qnm}) we see that 
\begin{equation}
 D^{QNM} = \frac{\sqrt{3} Q^3 \kappa}{4 M^2 R^2} .
\end{equation}
Since $D^{QNM}$ must be equal to $D$ evaluated in (\ref{hymode}) we see that 
we must have $\Phi_{12} =0$ for this system.

\section{Conclusions}

We have used the equilibrium partition function to obtain expressions for 7 parity 
odd transport coefficients which occur at 2nd order for a non-conformal fluid
with a single conserved charge. 
These transport coefficients can be expressed in terms of the anomaly, shear viscosity, bulk viscosity,
charge diffusivity and thermodynamic functions. 
Out of these $2$ transport coefficients can also be derived using the Kubo formulae. 
These formulae agree with that obtained by the partition function method. 
The equilibrium partition function also gives $3$ constraints for 5 other parity odd transport
coefficients at this order. 
The transport coefficient $\Phi_1$  affects chiral dispersion relations \cite{Kharzeev:2011ds}. 

As we have mentioned earlier, parity odd coefficients have be examined earlier for conformal 
charge transport 
in \cite{Kharzeev:2011ds}. There the principle used was that these coefficients should not contribute to 
entropy production. 
In general the constraints obtained by examining the zero entropy production condition
will be more than that obtained from the equilibrium partition function. 
It will be interesting to carry out the analysis of \cite{Kharzeev:2011ds} to non-conformal fluids and compare with  the results obtained in this paper. 

Our analysis of these transport properties were motivated by the possibility of studying them 
in holography.  While this work was being done we received preprint \cite{Megias:2013joa} which 
evaluates all 2nd order transport coefficients for a charged conformal fluid 
in the framework of AdS/CFT. 
It is  useful to compare  the relations we have obtained for the parity 
odd sector with the  expressions of \cite{Megias:2013joa}. 

Finally it will be useful to develop Kubo like expressions for all the parity odd transport 
coefficients. From the constitutive relations it seems that the remaining transport coefficients
involves $3$ point functions. Determining these relations will provide an alternate method to 
obtain the transport coefficients from holography.

\appendix 
\section{Classification of parity odd  data  at 2nd order in derivatives} \label{class}

In this appendix we provide some details  which led to the classification of 
the parity odd data at second order in derivatives given in table \ref{table:2ndorder}. 
We consider the following basis of vectors to construct the the second order 
terms: 
\begin{itemize}
\item
Parity odd vectors:
\begin{equation}
{\rm Vorticity:} \; l^\mu = \epsilon^{\mu\nu\alpha\beta}u_\nu \partial_\alpha u_\beta
\qquad {\rm Magnetic\; field:} \; B^\mu = \frac{1}{2} \epsilon^{\mu\nu\alpha\beta} u_\nu 
{\cal F}_{\alpha\beta}. 
\end{equation}
\item Parity even vectors
\begin{eqnarray}
u^\mu, \qquad  \partial_\mu T, \qquad \partial_\mu \nu, \\ \nonumber
{\rm Electric\; field:}\rightarrow V^\mu = \frac{E^\mu}{T} - P^{\mu \rho}\partial_\rho \nu.
\end{eqnarray}
\end{itemize}
Among these vectors, the electric field $V^\mu$ vanishes on  the equilibrium 
fluid configuration given in (\ref{back}). 
We also consider the shear tensor 
\begin{equation}
\sigma_{\mu\nu} = \nabla_{\langle\mu} u_{\nu\rangle}.
\end{equation} 
and the scalar 
\begin{equation}
\Theta = \nabla_\mu u^\mu. 
\end{equation}
Note that both $\sigma_{\mu\nu}$ and the scalar $\Theta$ vanishes on 
the equilibrium configuration in (\ref{back}). 
We use these basic quantities we can construct the various parity odd terms that occur at 
second order in derivatives given in table \ref{table:2ndorder}.  These terms are independent 
and cannot be related to each other by equations of motion to first order in derivatives.

Let us consider the scalars listed in table \ref{table:2ndorder}: One would have 
thought that scalars of the form $\nabla_\mu l^\mu$ and $ \nabla_\mu B^\mu$ should
be listed.  But it can be shown that these 
scalars can be related to the ones listed in the
table by   calculations which lead to the following identities.  
 \begin{eqnarray}\label{ex3}
  \nabla_\mu l^\mu &=& \left(\frac{2qT}{E + P}\right) V^\mu l_\mu - 2\left(\frac{\partial_\mu T}{T}\right)l^\mu\\
 \nabla_\mu B^\mu &=& \left(\frac{qT}{E + P}\right) V^\mu B_\mu -\left(\frac{\partial_\mu T}{T}\right)B^\mu - T l^\mu( V_\mu - \partial_\mu\nu)
  \end{eqnarray}
 These identities can be verified by straight forward calculations. 

Let us now consider the vectors listed in table \ref{table:2ndorder}: 
Note that the vectors ${\cal V}^{\mu}_{(3)}$ to ${\cal V}^{\mu}_{(9)}$ vanish
on the equilibrium configuration (\ref{back}). 
Again one would have naively expected to list terms such 
as $P^\mu_\alpha(u.\nabla) B^\alpha$. We will show now  
using the equations of motion to the zeroth order that this term is related to the 
vectors already listed in table \ref{table:2ndorder}. 
The  manipulations are as follows:
\begin{eqnarray}\label{eqn1a}
P^\mu_\alpha(u.\nabla) B^\alpha
&=& \frac{1}{2}P^\mu_\alpha u^\theta\nabla_\theta\left[\epsilon^{\alpha\nu\lambda\beta}u_\nu {\cal F}_{\lambda\beta}\right], \nonumber\\
&=&\frac{1}{2}P^\mu_\alpha \epsilon^{\alpha\nu\lambda\beta}{\mathfrak a}_\nu {\cal F}_{\lambda\beta} +\frac{1}{2}\epsilon^{\mu\nu\lambda\beta}u_\nu u^\theta \nabla_\theta {\cal F}_{\lambda\beta}, \nonumber\\
&=& -\epsilon^{\mu\nu\lambda\beta}{\mathfrak a}_\nu u_\lambda u^\theta{\cal F}_{\theta\beta} -\epsilon^{\mu\nu\lambda\beta}u_\nu u^\theta\nabla_\lambda {\cal F}_{\beta\theta}, \nonumber\\
&= &\epsilon^{\mu\nu\lambda\beta}{\mathfrak a}_\nu u_\lambda E_{\beta} -\epsilon^{\mu\nu\lambda\beta}u_\nu \nabla_\lambda E_\beta +\epsilon^{\mu\nu\lambda\beta}u_\nu (\nabla_\lambda u^\theta){\cal F}_{\beta\theta}, \nonumber \\
&=&-T\epsilon^{\mu\nu\alpha\beta}u_\nu \nabla_\alpha V_\beta  + \sigma^{\mu\nu}B_\nu  -\Theta\left[\frac{2B^\mu}{3}-T \left(s\frac{\partial\nu}{\partial s} +q\frac{\partial\nu}{\partial q}\right)l^\mu\right]\nonumber\\
&-&\left(\frac{qT^2}{E +P}\right)\epsilon^{\mu\nu\lambda\sigma}u_\nu V_\lambda \partial_\sigma \nu-\frac{1}{2}\epsilon^{\mu\nu\lambda\sigma}u_\nu l_{\lambda} B_{\sigma}, \nonumber\\
&=&-T\epsilon^{\mu\nu\alpha\beta}u_\nu \nabla_\alpha V_\beta  + {\cal V}_7^\mu-\frac{2{\cal V}_6^\mu}{3}+T \left(s\frac{\partial\nu}{\partial s} +q\frac{\partial\nu}{\partial q}\right) {\cal V}_3^\mu\nonumber\\
 &+& \left(\frac{qT^2}{E +P}\right) {\cal V}_{(9)}^\mu
+\frac{1}{2}{\cal V}_{(1)}^\mu. 
\end{eqnarray}
Here ${\mathfrak a}_\nu = (u.\nabla)u_\nu$. 
In the RHS of  equation (\ref{eqn1a}),  all terms except the last one vanishes at equilibrium.
Therefore  $(u.\nabla) B^\mu$ does not vanish in equilibrium but it can be 
related to all the other vectors listed in the table. 
A similar analysis can be done for the vector of the form 
 $P^{\mu}_{\alpha}(u.\nabla)l^{\alpha}$. We have 
\begin{eqnarray}\label{eqn2a}
P^\mu_\alpha(u.\nabla) l^\alpha &=& R\epsilon^{\mu\nu\alpha\beta}u_\nu \nabla_\alpha V_\beta  
 +\left(\frac{\partial R}{\partial T}\right)\epsilon^{\mu\nu\alpha\beta}u_\nu (\nabla_\alpha  T)V_\beta \\  \nonumber 
& & +\left(\frac{\partial R}{\partial \nu}\right)\epsilon^{\mu\nu\alpha\beta}u_\nu (\nabla_\alpha  \nu)V_\beta  +
\sigma^{\mu\nu}l_\nu+\left(\frac{s}{T}\frac{\partial\nu}{\partial s} +\frac{q}{T}\frac{\partial\nu}{\partial q}-\frac{2}{3}\right)\Theta l^\mu,
\end{eqnarray}
where 
\begin{equation}
R = \frac{q T}{E + P}.
\end{equation}
From equation (\ref{eqn2a}) it is clear that $P^\mu_\alpha(u.\nabla) l^\alpha$ vanishes at equilibrium
and that it is related to the other vectors listed in table  \ref{table:2ndorder}.

In fact using symmetries  we can list out all the independent terms appearing at the second order. 
We can show that  out at two derivatives either involving only the 
fluid variables or the velocity and derivative of the field strength, only one pseudo vector can be constructed if we demand on-shell independence for all the terms. This makes it clear that once we have chosen $\epsilon^{\mu\nu\alpha\beta}u_\nu \nabla_\alpha V_\beta$,
 all other two derivative pseudo vectors   of the form  mentioned earlier
 must be related to ${\cal V}^\mu_{(4)}$ by  equations of motion. From this argument 
 it is possible to observe  that it is not necessary to list 
  $(u.\nabla)B^\mu$, $(u.\nabla)l^\mu$, $\nabla_\mu l^\mu$ or $\nabla_\mu B^\mu$ as independent
  data.
  A similar analysis can be performed for the Pseudo-tensor. This leads to the 
  the $12$ tensors listed in \ref{table:2ndorder}.

\section{Consistency of the fluid profiles at the linear order}
\label{console}

In this section we show that the velocity perturbations
 and the background field configuration considered
in equations (\ref{bcke}) to  (\ref{velpert}) 
  consistently solve the linearized fluid equations of motion without the need of the 
any other fluctuations. 
The perturbations  depends only on the time, $t$ and the spatial direction, $y$. \\
We will consider the perturbations to the linear order.  We first  write down
all the components of the stress tensor to 2nd order in derivatives and to linear 
order in the perturbations. 
This are given by 
\begin{eqnarray}
\label{eqn3}
 T^{tt} &=& E^{(0)}, \\  \nonumber
T^{tx} &=& = ( E^{(0)} + P^{(0)} ) v^x + P^{(0)} h_{tx},  \\  \nonumber
T^{tz} & =&   ( E^{(0)} + P^{(0)} ) v^z + P^{(0)} h_{tz}, \\ \nonumber
T^{ty} &=&0, \\ \nonumber
T^{xx}  &=& P^{(0)}, \\  \nonumber
T^{xy} &=& -P^{(0)} h_{yx} - \eta( \partial_y v^x + \partial_t h_{yx}) \\ \nonumber
&& +\frac{1}{2}\Phi_1\partial_y^2 ( v^z + h_{zt} ) + \frac{1}{2} \Phi_2 \partial_y^2 a_z 
\\ \nonumber
& & + \frac{1}{2}\Phi_{12} \partial_y(  \partial_ y v^z  + \partial_t h_{yz} ), \\  \nonumber
T^{xz} & =& 0, \\  \nonumber
T^{yy} &=& P^{(0)}, \\  \nonumber
T^{yz}  &=& -P^{(0)} h_{yz} - \eta( \partial_y v^z + \partial_t h_{yz}) \\ \nonumber
&&-\frac{1}{2}\Phi_1\partial_y^2 ( v^x + h_{zt} ) - \frac{1}{2} \Phi_2 \partial_y^2 a_x  \\ \nonumber
& & - \frac{1}{2}\Phi_{12} \partial_y(  \partial_ y v^x  + \partial_t h_{y} ),  \\ \nonumber
T^{zz} &=& P^{(0)}. 
\end{eqnarray}
Now there are $4$ equations of motion for the stress tensor. We will show that 
that the $t$ and $y$ components are trivially satisfied. 
The $x$ and $z$ components of this equations can be used to determine the 
velocity profiles. 
The $t$ component of the equations of motion of the stress tensor is given by 
\begin{equation} \label{tteq}
 \partial_t T^{tt} + \partial_y T^{yt} + \Gamma^{\mu}_{\mu t} T^{tt} 
+ \Gamma^t_{tt} T^{tt} + \Gamma^{t}_{xx} T^{xx} 
+ \Gamma^t_{yy} T^{yy} + \Gamma^{t}_{zz} T^{zz} =0.  
\end{equation}
Note that we have used the fact that all Christofell symbols are first order in the 
fields. We have also used the fact that the only dependence is through time $t$ and $y$. 
Now examining the Christofell symbols given in  (\ref{eqn3.5}) we see that there is no contribution
to  the equation in (\ref{tteq}) 
from any  term involving the Christofell symbols. 
 Also since $T^{tt} = E^{(0)}$, the first term also vanishes. 
The second term in the equation vanishes because $T^{yt}=0$. 
Thus this equation is satisfied and imposes no constrains on the velocity configuration
chosen. 
Now let examine the $y$ component of the equations of motion of the stress tensor. 
We have 
\begin{equation}
 \partial_t T^{ty} + \partial_y T^{yy}   +\Gamma^{\mu}_{\mu y} T^{yy}
+ \Gamma^y_{tt} T^{tt} + \Gamma^{y}_{xx} T^{xx} 
+ \Gamma^y_{yy} T^{yy} + \Gamma^{y}_{zz} T^{zz} =0.  
\end{equation}
Again, examining each term one can see that this equation is also satisfied. 
Thus the $t$ and $y$ equations are satisfied. 
Thus $v_x$ and $v_y$ are determined by the $x$ and $z$ component of the 
equations of motion
\begin{eqnarray}
\partial_t T^{tx} + \partial_t h_{tx} T^{tt} + \partial_y T^{yx} =0, \\ \nonumber
\partial_t T^{tz} + \partial_t h_{tz} T^{tt} + \partial_y T^{yz} =0. 
\end{eqnarray}

To complete the analysis we show that the equations of motion of the 
charge current are also trivially satisfied. 
The only non-zero values of the the vorticity to the linear order is given by 
\begin{equation}
 l^x = \partial_y v_z, \qquad l^z =- \partial_y v_x.
\end{equation}
Similarly the non-zero values of the magnetic field to the linear order 
\begin{equation}
 B^x = \frac{1}{2} \partial_y a_z, \qquad B^z = - \frac{1}{2} \partial_y a_x.
\end{equation}
One can also see that for the background in (\ref{bcke}) and fluctuations  to the linear order the 
electric field $V^\mu$ vanishes to the linear order in the fields.
It can also be seen that  to the linear order in fields the vectors
${\cal V}^\mu_{(i)}$  with $i=1, \cdots 9$ all vanish.  
Now using all this information, the currents to the linear order in 
fields and to second order in derivatives are given by 
\begin{eqnarray}
 J^t &=& -q^{(0)} , \\ \nonumber
J^x &=& q^{(0)} v^x + \xi l^x + \xi_B B^x, \\ \nonumber
J^y &=& 0,  \\ \nonumber
J^z &=& q^{(0)} v^z  +  \xi l^z + \xi_B B^z. 
\end{eqnarray}
Now the current conservation equation to this order reads
\begin{equation}
 \partial_t ( \sqrt{g} J^t) + \partial_y ( \sqrt{g} J^y) =0, 
\end{equation}
and since $\sqrt{g}$ does not change to the linear order, this equation is 
satisfied and does not impose any further conditions on the velocities 
$v^x$ and  $v^y$.  Here we have used that the only dependence is 
through $t$ and $y$. 

Thus the constant background with  the linear velocity profiles $v^x, v^y$ 
given in (\ref{bcke}) to (\ref{velpert})  consistently solve the equations of motion.  
There are only $2$ equations which determine the velocity profiles $v^x, v^y$. 
A simple way of stating this is that we have turned only  the shear fluctuations which 
decouple from  the rest of the modes. \\ \\

\acknowledgments
 We thank Aninda Sinha for organizing  a stimulating 
workshop on ``Non-perturbative gauge theories, holography and all that'' 
which initiated this collaboration. 
We thank Nabamita Banarjee, Bobby Ezuthachan,  Yashodhan Hatwalne, Sachin Jain, Shiraz
Minwalla  and  Rahul Pandit for useful and stimulating discussions.
S.B thanks the CHEP, IISc and the ICTS-TIFR for hospitality during the early phase of this 
work.  The work of J.R.D is partially supported by  the Ramanujan fellowship
DST-SR/S2/RJN-59/2009.

%\bibliography{parity2}
%\bibliographystyle{JHEP}

\providecommand{\href}[2]{#2}\begingroup\raggedright\endgroup

\end{document}